%% file: igmpaper.tex
\documentclass[portrait]{emulateapj}

\usepackage[squaren]{SIunits}

\usepackage{ wasysym }
\usepackage{graphicx}
\usepackage{epstopdf}

\slugcomment{Sumitted to The Astrophysical Journal.}

\shorttitle{The Low-$z$ IGM}
\shortauthors{Tilton et al.}

\begin{document}

\title{The Low-Redshift Intergalactic Medium as Seen in Archival Legacy Hubble/STIS and FUSE Data}

\author{Evan M. Tilton, Charles W. Danforth and J. Michael Shull}
\affil{CASA, Department of Astrophysical and Planetary Sciences, University of Colorado, 389-UCB, Boulder, CO 80309; {evan.tilton@colorado.edu, charles.danforth@colorado.edu, michael.shull@colorado.edu}
}
\and
\author{Teresa L. Ross}
\affil{Department of Astronomy, New Mexico State University, Las Cruces, NM 88003; {rosst@nmsu.edu}
}

\begin{abstract}
We present a comprehensive catalog of ultraviolet (\textit{HST}/STIS and \textit{FUSE}) absorbers in the low-redshift IGM at $z<0.4$. The catalog draws from the extensive literature on IGM absorption, and it reconciles discrepancies among previous catalogs through a critical evaluation of all reported absorption features in light of new \textit{HST}/COS data. We report on 746 \ion{H}{1} absorbers down to a rest-frame equivalent width of \unit{12}{\milli\angstrom} over a maximum redshift path length $\Delta z=5.38$. We also confirm 111 \ion{O}{6} absorbers, 29 \ion{C}{4} absorbers, and numerous absorption features due to other metal ions. We characterize the distribution of absorber line frequency as a function of column density as a power law, $d\mathcal{N}/dz\propto N^{-\beta}$, where $\beta=2.08\pm0.12$ for \ion{O}{6} and $\beta= 1.68\pm0.03$ for \ion{H}{1}. Utilizing a more sophisticated accounting technique than past work, the catalog accounts for $\sim43\%$ of the baryons: $24\pm2\%$ in the photoionized Ly$\alpha$ forest and $19\pm2\%$ in the WHIM as traced by \ion{O}{6}. We discuss the large systematic effects of various assumed metallicities and ionization states on these calculations, and we implement recent simulation results in our estimates. 
\end{abstract}



\section{Introduction}
The successful connection of recent advances in cosmology to the processes that shape the low-redshift universe requires a detailed understanding of the intergalactic medium (IGM). Simulations \citep[e.g.,][]{cen99, dav01,smi11} predict that a large portion of the baryonic mass of the universe resides in the non-luminous IGM in the form of primordial gas and gas processed by galaxies. The systematic characterization of these systems in absorption at different redshifts is thus a key step toward understanding the evolution of galaxies and the universe in general.

At high-redshift, the diffuse photoionized IGM, commonly known as the Ly$\alpha$ forest, accounts for nearly all of the baryonic matter, but simulations and observations both find that this fraction is reduced to $\sim30\%$ by $z\sim0$ \citep[e.g.,][]{dav01,pen00}. Many of the remaining baryons reside in the ``warm-hot'' IGM (WHIM) phase, which is heated to $\unit{10^{5-7}}{\kelvin}$ as gas is shocked during gravitational infall, cloud-cloud collisions, and by galaxy winds \citep{cen06}. Because of their low density, these phases of the IGM are most readily observed in absorption against a background continuum source such as an AGN, using Lyman series \ion{H}{1} lines as well as highly ionized metal species such as \ion{C}{4} and \ion{O}{6}. As these features lie in the rest-frame far-ultraviolet (FUV), recent low redshift ($z\lesssim 0.4$) studies \citep[e.g.,][]{dan06,dan05,dan08,leh07,tho08,tri08} have focused on observations with the \textit{Far Ultraviolet Spectroscopic Explorer} (\textit{FUSE}) and the Space Telescope Imaging Spectrograph (STIS) on the \textit{Hubble Space Telescope} (\textit{HST}). Together, these studies suggest that the column density distribution of \ion{H}{1} follows a power law, $d\mathcal{N}/dN\propto N^{-\beta}$, where $\beta\sim 1.7$, and that highly ionized metals such as \ion{O}{6} follow a similar distribution with $\beta\sim 2$.

With the installation of the Cosmic Origins Spectrograph (COS) on \textit{HST}, our sensitivity in the FUV is at an all-time high, and we have the opportunity to probe the IGM more deeply than ever before. With more than ten times the sensitivity of STIS , COS \citep{gre11} promises to probe the IGM with a larger and more systematic selection of sight lines down to lower column densities. Already, a large number of targets have been observed \citep[e.g.,][]{dan10,dan11,nar11,sav11}. A systematic understanding of existing STIS and \textit{FUSE} data in light of the early COS data, however, is essential to guiding future IGM studies. To this end, we present in this work a systematic catalog of STIS and \textit{FUSE} IGM absorbers, the largest such catalog to date, in light of past studies and new COS data. We draw primarily from the extensive literature in this field, creating a comprehensive database of IGM absorbers that reconciles discrepancies among previous lists of absorbers. It contains 746 \ion{H}{1}  and 111 \ion{O}{6} absorbers, as well as detections of many other ion species, along 44 AGN sight-lines. We further present new counting statistics and cosmological parameters derived from the catalog, and we compare these results to other work in the field. This catalog will lay the groundwork for future COS IGM studies.
\newpage

\section{Methodology}

\subsection{Sight-Line Selection and Data Reduction}

The primary goal of this catalog is to create the most comprehensive low-redshift IGM census possible with legacy STIS and \textit{FUSE} data. Therefore, we included all 44 sight lines with  $z_{AGN}\lesssim 0.5$ for which STIS/E140M and/or \textit{FUSE} data were available at reasonable quality. The imposed high-redshift cutoff guarantees STIS wavelength coverage of nearly all potential Ly$\alpha$ features while avoiding complications in interpretation due to high densities of IGM absorption features present in higher-redshift sight lines ($z\sim 1-2$). All data have a resolution of \unit{20}{\kilo\meter\usk\reciprocal\second}  or better at a signal-to-noise ratio per resolution element $\mathrm{(S/N)}_{\mathrm{res}}\gtrsim 5$ in most cases. The \textit{FUSE} data provide wavelength coverage over \unit{905-1187}{\angstrom}, while, in most cases, STIS/E140M covers \unit{1162-1729}{\angstrom}. A few of the sight-lines (3C 273, Akn 564, Mrk 509, Mrk 205, and PG 1116+215) lack the longest wavelength order of the STIS/E140M data and truncate at \unit{1709}{\angstrom}. The data were supplemented by COS data in a few cases to assist in the confirmation and interpretation of absorption features as described in Section~\ref{sec:construct}. These COS data add additional coverage over \unit{1135-1770}{\angstrom} at a resolution of $\sim\unit{15}{\kilo\meter\usk\reciprocal\second}$ and a S/N that greatly improves upon the \textit{FUSE} and STIS data in most cases. We used all COS data available for these sight lines as of 2011 September 28. Table~\ref{tab:sightlines} summarizes the final sample of sight lines and data, which cover a total Ly$\alpha$ redshift pathlength of $\Delta z=5.38$.\input{ovi_sightlines}

\defcitealias{dan08}{DS08}

The STIS data were reduced as described in our group's previous paper on this subject \citep[hereafter DS08]{dan08}, and the \textit{FUSE} data were reduced as described by \citet{dan06}. We relied primarily on the highest throughput \textit{FUSE} channel, LiF1 \citep{moo00}, but we also used the LiF2 and SiC channels to aid in interpretation and to provide coverage where the LiF1 data were absent, suspect, or of low S/N. The COS data were reduced as described in \citet{dan10}.

\subsection{Catalog Construction}
\label{sec:construct}
\begin{deluxetable*}{rr||rr}
\tabletypesize{\tiny}
\tablecaption{Sources of data used in the catalog\label{tab:sources}}
\tablewidth{0pt}
\tablehead{
\colhead{Number\tablenotemark{a}} & \colhead{Reference} & \colhead{Number\tablenotemark{a}} & \colhead{Reference}
}
\startdata
1 & \citet{wil06} & 12 & \citet{sem04} \\
2 & \citet{dan08} & 13 & \citet{ric04} \\
3 & \citet{tho08} & 14 & \citet{jen03} \\
4 & \citet{pro04} & 15 & \citet{nar11} \\
5 & \citet{leh07} & 16 & \citet[Table 3]{tri08} \\
6 & \citet[Table 2]{tri08} & 17 & \citet{sav05} \\
7 & \citet{sem01} & 18 & \citet{dan06} \\
8 & \citet{tri01} & 19 & \citet{how09} \\
9 & \citet{oeg00} & 20 & \citet{ric01} \\
10 & \citet{ara06} & 21 & \citet{tum05} \\
11 & \citet{sav02} & 22 & \citet{leh06} \\

\enddata
\tablenotetext{a}{Identifying number used in the catalog.}
\end{deluxetable*}
The creation of the catalog was designed as a critical analysis of previously published lists of absorption features in the low-redshift IGM. We began by correlating the features reported in a variety of sources, summarized in Table~\ref{tab:sources}, for all sight lines in our sample. Although some of the sight lines were analyzed in only one previous work, many have been studied by two or more groups. One sight line, Ton S210, has data from \textit{FUSE}, STIS, and COS, but it has not been included in any major, multi-ion IGM absorption survey. As such, our analysis of that sight line is original. Each feature was individually inspected in the data, and its interpretation, in terms of both structure and identification, was confirmed in context with related lines. We used any available COS data to aid in the interpretation of each feature, but the measurements themselves were not made from the COS data. Although the COS data sometimes refute the existence of features that are seen by the other instruments, the features were nevertheless included in the catalog so as to maintain a homogeneous sample; they were, however, flagged as suspect. For cases in which our interpretation and all existing literature measurements agreed in the basic interpretation of the feature, we computed consensus values for the observed wavelength, rest-frame equivalent width and Doppler $b$-values as means of the measurements reported in the literature. One-sigma errors were computed assuming that the measurements were independent. This assumption, while not formally accurate, is roughly approximated by differences in data reduction, continuum placement, and measurement technique. As a complementary measure of the error, we also computed the mean absolute deviation among the individual measurements. In cases where one or more of the sources disagreed on the interpretation of the feature, we critically evaluated the arguments and data and either chose the best interpretation or, if necessary, remeasured the feature before proceeding as in the case of total agreement. If appropriate, alternate identifications and structural interpretations were noted. Throughout this process, a number of previously unreported features were found serendipitously. Although we measured them and included them in the catalog, we note that this project was not a systematic search for new absorption systems, owing to its literature-based nature. Every redshift with an \ion{H}{1} absorption system, however, was checked for corresponding absorption in each of the metal ions listed in Table~\ref{tab:ions}.\begin{deluxetable}{lrlccc}
\tabletypesize{\tiny}
\tablecaption{Primary IGM Diagnostic Lines\label{tab:ions}}
\tablewidth{0pt}
\tablehead{
\colhead{Ion} & \colhead{$\lambda_{\rm rest}$} & \colhead{$f$\tablenotemark{a}} & \colhead{$[X/H]_{\astrosun}$\tablenotemark{b}}  & \colhead{$f_{\rm ion}$\tablenotemark{c}} & \colhead{$\log T_{\rm max}$}\\
 & \colhead{(\AA)} & & & \colhead{CIE}&
}
\startdata
H\,I & 1215.67 & 0.4164 & 0.00 & &$<4.19$\tablenotemark{d} \\
 & 1025.72 & 0.07914 & 0.00  & &$<4.19$\tablenotemark{d} \\
Fe\,II & 1144.94  &0.0830 & -4.50  &0.79&$4.05-4.27$\tablenotemark{d} \\
 & 1063.18 &0.0547 & -4.50  &0.79&$4.05-4.27$\tablenotemark{d} \\

C\,II & 1334.53 & 0.128 & -3.57  &0.97 &$4.15-4.66$\tablenotemark{d} \\
 & 1036.34  &0.118  & -3.57  & 0.97&$4.15-4.66$\tablenotemark{d} \\
Si\,II & 1260.42  &1.18 & -4.49  &0.97&$<4.30$\tablenotemark{d} \\
 &1193.29  &0.582 & -4.49  &0.97&$<4.30$\tablenotemark{d} \\
S\,II & 1259.52 &0.0166 & -4.88  &1.00&$<4.47$\tablenotemark{d} \\
& 1253.81 &0.0109 & -4.88  &1.00&$<4.47$\tablenotemark{d} \\

Fe\,III & 1122.52 & 0.0544 & -4.50  & 0.89&$4.27-4.58$\tablenotemark{d} \\
Si\,III & 1206.50 & 1.63 & -4.49  &0.90 &$4.30-4.85$\tablenotemark{d} \\
C\,III & 977.02 & 0.757 & -3.57  &0.83 &$4.66-5.01$\tablenotemark{d} \\
S\,III &1012.50  &0.0438 & -4.88  &0.84&$4.47-4.93$\tablenotemark{d} \\
 & 1190.20 &0.0237 & -4.88  &0.84&$4.47-4.93$\tablenotemark{d} \\

Si\,IV & 1393.76 & 0.513 & -4.49  &0.35 &$4.8$\tablenotemark{e} \\
 & 1402.70 & 0.254 & -4.49 & 0.35 &$4.8$\tablenotemark{e} \\
C\,IV & 1548.20 & 0.190 & -3.57 &0.29 &$5.0$\tablenotemark{e} \\
 & 1550.77 & 0.0948 & -3.57  &0.29 &$5.0$\tablenotemark{e} \\
S\,IV & 1062.66 &0.0494 & -4.88  &0.61&$4.93-5.13$\tablenotemark{d} \\

N\,V & 1238.82 & 0.156 & -4.17  &0.24 &$5.25$\tablenotemark{e} \\
 & 1242.80 & 0.0777 & -4.17 &0.24 &$5.25$\tablenotemark{e} \\
O\,VI & 1031.93 &0.1325 & -3.31 &0.22 &$5.45$\tablenotemark{e} \\
 & 1037.62 &0.0658 & -3.31 &0.22 &$5.45$\tablenotemark{e} 

\enddata
\tablenotetext{a}{{Absorption line oscillator strength \citep{mor03}.}}
\tablenotetext{b}{Solar photospheric element abundance, $\log(X/H)$ \citep{asp05}.}
\tablenotetext{c}{Peak ion fraction under CIE \citep{sut93}.}
\tablenotetext{d}{Temperature range ion is dominant under CIE \citep{sut93}.}
\tablenotetext{e}{Ion is never dominant under CIE; temperature is peak CIE abundance.}
\end{deluxetable}

\defcitealias{nistasd}{NIST Atomic Spectra Database}

When we measured a feature's properties, we used either the apparent optical depth (AOD) method \citep{sem92} or a Voigt profile fit \citep[see discussion in][and sources therein]{dan06}. We qualitatively judged which method was most appropriate for each feature, taking into account the two methods' relative strengths. Although the two methods usually generate comparable results, the AOD method performs better for weak lines or in extremely noisy regions of the spectra, while profile fitting is preferable for strong or blended lines or lines with complicated component structure \citepalias{dan08}. Where necessary, we adopted the atomic parameters given by \citet{mor03} or, for lines not included in that work, from the \citetalias{nistasd} \citep{nistasd}.

We took a conservative approach to the splitting of absorption features into separate components. In cases where clear separation was visible or multiple lines of the same absorber showed consistent asymmetry, we favored multiple-component interpretations. On the other hand, if separation into components was ambiguous, we favored a single-component interpretation, even if the overall profile shape suggested the presence of multiple components. Where possible, we used high-S/N COS data to further inform our interpretation of the component structure. This approach leaves the catalog with few, if any, false positives in terms of measured components, but some features may have more components than reported here. For comparison, the features reported in this catalog yield counting statistics intermediate between those of \citetalias{dan08}, which generally used single-component interpretations, and the profile-fitting results of \citet{tri08}, who generally used many-component interpretations. Our component splitting is comparable to that of \citet{tho08}. Differences in component-splitting approaches are expected to have little impact on the calculation of baryon densities because the primary methods used to derive those quantities depend only on the total column density observed.

After compiling complete lists of the properties of individual absorption lines from a given system, we took a similar approach in determining the consensus properties of the absorber inferred from those lines. In cases where only a single line was present for the absorber, the Doppler $b$-value was taken to be the same as the mean value for the feature. The column density was computed as the geometric mean of the literature values, with errors treated as above. In many cases, more than one absorption line was present for the absorber. For metal species with  multiple lines, we proceeded as with single-line detections but used a mean weighted by the expected relative strengths of the observed absorption lines. If any of the lines was deemed suspect, for example because of excessive noise or blending, it was not included in the mean. The \ion{H}{1} absorbers were treated in a more complicated fashion, because, in addition to allowing measurements of the individual Lyman series lines, the ion is particularly conducive to curve-of-growth (COG) analyses \citep[see discussion in][]{dan06} when multiple lines are present. In cases of agreement among the various literature sources, we were able to compute simple mean values for the column density and $b$-value of the absorber as above. However, if any remeasurement or reinterpretation was necessary at the level of individual absorption lines, we used either new COG analyses or new weighted means of the line measurements to determine the Doppler $b$-values and column densities for the absorber. In these consensus mean at the absorber level, we also included measurements from literature sources that did not report properties for the individual absorption lines. In a few cases with extremely high column density \ion{H}{1}, we estimated the absorber parameters by fitting the integrated absorptin from higher Lyman lines (i.e., the Lyman decrement), if sufficiently short-wavelength data were available. The vast majority of \ion{H}{1} absorbers, however, show only Ly$\alpha$ absorption, and in those cases the absorber-level properties were inferred directly from that single line.

Because the various literature sources differ slightly in their approach to reporting their measurements, we made minor adjustments to our general approach in individual cases. \citet{tri08} report both total equivalent width and column density for each absorption complex as well as profile-fitting measurements of individual components; we used the relevant measurements as appropriate to our interpretation of the structure of the feature. In using the measurements of the \citet{leh06} study of the HE 0226-4110 sight line, we adopted the AOD equivalent width and Doppler $b$-value measurements and the profile-fitting column density measurements. The \citet{wil06} study of the PKS 0405-123 sight line employed a methodological approach that resulted in a large number of very weak \ion{H}{1} detections which, from an examination of the COS data, seem to be noise features rather than real absorption features. While we have included these measurements in the catalog for the sake of completeness, their equivalent widths and column densities have been flagged as upper limits to avoid biasing the sample set. Similarly, the \citet{ric04} study of the PG 1259+593 sight line reports a large number of ``probable'' Ly$\alpha$ detections. All were re-evaluated in light of the much higher S/N COS data and included in our catalog, but, as with the \citet{wil06} paper, we flagged those that seem be noise as upper limits rather than measurements. Finally, \citet{ric04} report a variety of measurements for each detected feature. For metal features, we favored the profile-fitting measurements where available and otherwise used the AOD measurements. For \ion{H}{1}, we used the profile-fitting measurements for individual features and, when possible, the COG measurements for ion-level absorber properties.

A number of features in the catalog are reported as upper limits, owing to excessive blending, weakness relative to noise, or for the reasons described in the previous paragraph. In these cases, we reported the most conservative limit from the references, which may be either a measured value in the case of blended lines or a limit based on the noise characteristics of the data. Because upper limits are treated somewhat differently among diffferent authors, we report a minimum S/N-based equivalent width for each feature for the sake of consistency:
\begin{eqnarray}
\label{eqn:wmin}
W_{\rm min}=\frac{4\lambda}{7200({\rm S/N})_{\lambda}}(1+z)^{-1}.
\end{eqnarray}
This quantity roughly corresponds to a 4 $\sigma$ significance level in equivalent width for the data in the rest frame of the absorber. It assumes a resolved feature with $b=\unit{25}{\kilo\meter\usk\reciprocal\second}$, roughly the median $b$-value for \ion{H}{1} and \ion{O}{6}, with the factor of $7200$ corresponding to the resolution, $R=\lambda /\Delta\lambda$, required to resolve a line with that $b$-value. The signal-to-noise, $(\rm S/N)$, was calculated as the ratio of mean flux to its standard deviation in the smoothed continuum. Features with equivalent widths near or below the reported value of $W_{\rm min}$ should be viewed with skepticism and treated as upper limits. Note, however, that these 4 $\sigma$ estimates do not account for fixed-pattern noise, which is mitigated by the smoothing applied during the data reduction, or for other systematic effects.

Table~\ref{tab:linestab} presents the complete catalog of individual absorption lines. Columns (1) through (4) identify the feature according to sight line, observed wavelength, redshift, and transition, respectively. Column (5) lists an ``ion code'' that correlates absorption lines within a sight line in Table~\ref{tab:linestab} with the associated absorber properties in Table~\ref{tab:ionstab}. Columns (6) through (9) describe the measured strength of the feature. Column (6) is a flag that identifies the rest-frame equivalent width listed in Column (7) as a normal measurement (``n''), an upper limit (``\textless ''), a lower limit (``\textgreater ''), or an order-of-magnitude estimate (``  {\raise.17ex\hbox{$\scriptstyle\sim$}}'') when a reliable measurement could not be obtained. Column (8) is the mean absolute deviation of the measurements used to generate that consensus value. Column (9) identifies the sources or techniques used to construct the consensus value, with every two-digit portion of the value in the table corresponding to a source listed in Table~\ref{tab:sources}. Other numbers are used to identify our own measurements: 23 corresponds to a Voigt fit, 24 corresponds to an AOD measurement, 25 corresponds to a COG measurement, and 26 corresponds to a Lyman decrement estimate. Columns (10) through (13) report on the velocity width of the feature in terms of the Doppler $b$-parameter in the same format in which the equivalent width measurements were reported. Column (14) lists the quantity $W_{\rm min}$ as defined in Equation~\ref{eqn:wmin} and evaluated at the observed wavelength. Column (15) lists possible alternate line identifications in the rare case that identification was ambiguous. Finally, Column (16) flags inconsistencies between the COS data (where available) and the STIS and/or FUSE data. No flag is present if COS data are absent or in agreement with the other data, while flags are listed for features that are entirely absent from the COS data (``a''), weaker or partially absent in the COS data (``p''), or that display a different shape or structure in the COS data (``d'').
\input{linestab}
\input{ionstab}Table~\ref{tab:ionstab} presents the complete catalog of absorbers. Columns (1) through (3) identify the absorber via the sight line, redshift, and ion species, respectively. Column (4) reports the ion code that correlates Tables~\ref{tab:linestab} and \ref{tab:ionstab}, as described in the the previous paragraph. Using a similar format to the measurements in Table~\ref{tab:linestab}, Columns (5) through (7) report on the column density of the absorber, with Column (5) identifying the type of measurement with the same set of flags used in Table~\ref{tab:ionstab}, Column (6) reporting the column density, and Column (7) reporting the references or techniques used to generate the measurement, again as in Table~\ref{tab:linestab}. Columns (8) through (10) report on the measurements of the Doppler $b$-parameter in the same manner.

\subsection{Absorber Statistics}
\input{tablepaper}
In compiling absorber statistics, we followed the approach of \citetalias{dan08}. We focus on fourteen ions, listed in Table~\ref{tab:ions} along with some of their properties. For all statistics and derived quantities, we included only absorbers in the redshift range $\unit{500}{\kilo\meter\usk\reciprocal\second} < cz <(cz_{\rm AGN}-\unit{1500}{\kilo\meter\usk\reciprocal\second})$ and further required that $z<0.4$. These limits help to exclude absorbers intrinsic to either the Milky Way or the background AGN while also ensuring that the sample is restricted to the local, low-redshift universe where sensitive \ion{H}{1} measurements via the Ly$\alpha$ transition can be obtained. We further excluded all absorbers at $z>0.104$ in the Q1230+115 sight line ($z_{\rm AGN}=0.1170$) because these features are likely intrinsic to the AGN \citep{gan03}. Despite these velocity limits, some high-velocity AGN outflow systems may contribute to our sample. These limits are less restrictive than those adopted by \citet{tri08} and comparable to those adopted by \citetalias{dan08}.

For each ion species, we constructed simple histograms of the number of absorbers in 0.2 dex bins in $\log N$, as well as distributed histograms in which each detection is characterized by a Gaussian \citepalias{dan08}. We proceeded to calculate the number of absorbers  per unit redshift, 
\begin{eqnarray}
\frac{d\mathcal{N}(\log N)}{dz}=\sum_{i}\frac{\mathcal{N}_i (\log N_i)}{\Delta z_i}.
\end{eqnarray}
This quantity is a function of column density, $N$, because it must be corrected for incompleteness. The correction is acheived by dividing the number, $\mathcal{N}_i$, per bin by an effective redshift path length, $\Delta z_i$, which is itself a function of column density and accounts for variations in sensitivity to the absorption feature in question.

We establish this redshift pathlength by considering the S/N in the continuum of the highest-S/N data covering a given wavelength. This $(\rm S/N)_\lambda$ vector was rebinned to a resolution comparable to the data. The smoothed, normalized flux was then used to mask $(\rm S/N)_\lambda$ such that areas of saturated absorption were set to $(\rm S/N)_\lambda=0$ while regions of unsaturated absorption were scaled with the optical depth, $\tau_\lambda$, as $(\rm S/N)_\lambda\propto (\tau_\lambda +1)^{-1}$. This approach effectively removes lines of any type from the effective pathlength. The resulting vector was converted to a 4 $\sigma$ minumum equivalent width vector defined, as in \citetalias{dan08}, by
\begin{eqnarray}
W_{\rm min}=\frac{4\lambda}{R({\rm S/N})_{\lambda}}
\end{eqnarray}
where the instrumental resolution, $R=\lambda/ \Delta \lambda$, was set to 15,000 or 42,000 for \textit{FUSE} or STIS data, respectively. The effective redshift pathlength for singlet lines is then simply the total path length over all sightlines such that $W_\lambda >W_{\rm min}$. This is easily extended to multiplet lines  of different line strengths ($f\lambda$) by requiring, in the rest frame of the weaker line, that either $W_\lambda >W_{\rm min}(z_1)$ or $W_\lambda >[W_{\rm min}(z_2)](f_1\lambda_1/f_2\lambda_2)$.

Uncertainty in $d\mathcal{N}/dz$ is dominated by the uncertainty in absorber number ($\mathcal{N}_i$) per bin due to Poisson statistics \citep{geh86}. Inherent uncertainty in effective redshift pathlength $\Delta z_i$ is negligible, but this quantity contributes uncertainty arising from its variation over the width of a column density bin. We approximate this uncertainty as 
\begin{eqnarray}
d(\Delta z_i)=\frac{1}{8}\left [ (\Delta z)_{i+1}- (\Delta z)_{i-1} \right ],
\end{eqnarray}
as suggested by \citetalias{dan08}, in which these two sources of error were added in quadrature.
\newpage

\subsection{Baryon Fraction}
Studies of the cosmic microwave background \citep{kom11} have provided accurate measurements of the cosmological baryon density, $\Omega_b=0.0455\pm0.0028$, as a fraction of the critical density of the universe, $\rho_{\rm cr}=(3H_0^2/8\pi G)=(9.205\times 10^{-30}~ {\rm g~ cm^{-3}})h_{70}^2$ for a Hubble constant, $H_0=(70~{\rm km~s^{-1}~Mpc^{-1}})h_{70}$. The majority of this matter, both at high and low redshift, must reside in the IGM. A complete accounting of IGM baryons at low redshift is therefore best studied with a combination of FUV and X-ray spectroscopy \citep{shu12}. We address the former here with our catalog, which traces IGM gas at $T \lesssim 10^6~ \rm K$, by considering \ion{H}{1} absorbers in the Ly$\alpha$ forest and metal species as tracers of the WHIM.

\subsubsection{Metal Ions}

We present two quantities that measure baryon density contributions as traced by metal ions. In both cases, we use five equally spaced bins in $z$, and we assume a mean atomic mass, $\mu=1.32m_{\rm H}$, for ${ Y_{\rm He}}=0.2477$ \citep{pei07}. The contribution to the critical density by the IGM mass of an ion is given by\bigskip
\begin{eqnarray}
\Omega_{\rm ion}&=&\frac{H_0m_{\rm ion}}{c \rho_{\rm cr}}\nonumber \\
&\times& \int^{z_{\rm max}}_{0} \int^{N_{\rm max}}_{N_{\rm min}}  \frac{d\mathcal{N}(\log N)}{dz} N d(\log N)dz \nonumber \\
&=&(\unit{1.365\times 10^{-23}}{\centi\meter\squared})h_{70}^{-1}\left ( {m_{\rm ion}}/{\rm amu}\right ) \nonumber\\
& \times& ~\sum_{j=0}^{z_{\rm max}}~\sum_{i=\log N_{\rm min}}^{\log N_{\rm max}}\left ( \left [ \frac{d\mathcal{N}(\log N)}{dz}\right ]_{i,j} \langle N\rangle_{i,j} \right. \nonumber \\
&&~~~~\left. \times~\Delta\log N\frac{\Delta z_j}{z_{\rm max}}\right ) ,
\end{eqnarray}
where $m_{\rm ion}$ is the ion mass and the integration is over column density (\unit{\centi\meter\rpsquared}) and redshift path length. We report the uncertainty in $\Omega_{\rm ion}$ as the asymmetric Poisson uncertainty in $d\mathcal{N}$ evaluated at  the mean $dz$, $\langle z \rangle$, and $\langle N \rangle$ weighted by $d\mathcal{N}$. 

While $\Omega_{\rm ion}$ is the observed contribution to the critical density by a given ion, we are also interested in estimating the total baryonic density of all species as traced by a particular ion. We denote this estimate as $\Omega^{\rm (ion)}_{\rm IGM}$; it is obtained by correcting $\Omega_{\rm ion}$ for metallicity and ionization fraction. Generally, we assume a metallicity of $Z=0.1Z_{\odot}$ \citepalias{dan08} and an ionization fraction equal to the peak collisional ionization equilibrium (CIE) ion abundance, $f_{\rm ion}$, listed in Table~\ref{tab:ions}, but we discuss alternate choices for the product $Zf_{\rm ion}$ in Section~\ref{sec:baryon}. The quantity $\Omega^{\rm (ion)}_{\rm IGM}$ is thus
\begin{eqnarray}
\Omega^{\rm (ion)}_{\rm IGM} &=& \left ( \frac{H_0}{c \rho_{\rm cr}}\right )^2 \frac{\mu m_H}{Z({\rm M/H})_{\odot}f_{\rm ion}} \nonumber\\
&\times&~\int^{z_{\rm max}}_{0}\int^{N_{\rm max}}_{N_{\rm min}} \frac{d\mathcal{N}(\log N_{\rm ion})}{dz}\langle N_{\rm ion}\rangle d(\log N_{\rm ion})dz \nonumber \\
&=& \frac{1.83\times 10^{-23}~h_{70}^{-1}~{\rm cm^2}}{Z({\rm M/H})_{\odot}f_{\rm ion}} \nonumber \\
&\times&~\sum_{j=0}^{z_{\rm max}}~\sum_{i=\log N_{\rm min}}^{\log N_{\rm max}}\left ( \left [ \frac{d\mathcal{N}(\log N_{\rm ion})}{dz}\right ]_{i,j} \langle N_{\rm ion}\rangle_{i,j} \right. \nonumber \\
&&~~~~\times~\left. \Delta\log N_{\rm ion}\frac{\Delta z_j}{z_{\rm max}}\right ),
\end{eqnarray}
where the factor of ${\Delta z_j}/{z_{\rm max}}$ weights the contribution of each bin in $dz$. Uncertainties in $\Omega^{\rm (ion)}_{\rm IGM}$ are calculated as for  $\Omega_{\rm ion}$; systematic uncertainties arising from our assumptions are not included. Additional systematic uncertainties in this quantity likely exist due to effects of nonequilibrium ionization, photoionization, or deviations from the assumed $10\%$ solar metallicity. \citet{shu12} looked into these effects in detail, especially with respect to \ion{O}{6}, and we discuss them further in Section~\ref{sec:baryon}

\subsubsection{The Ly$\alpha$ Forest}
\label{sec:omega}
The Ly$\alpha$ forest is an important resevoir of baryons, with previous surveys finding up to $\sim30\%$ the total baryonic content of the universe in this photoionized phase of the IGM at low redshift \citep[e.g.,][]{pen00, leh07, dan08}. As the catalog presented here contains the largest sample of Ly$\alpha$ forest absorbers to date, with a total Ly$\alpha$ redshift pathlength $\Delta z=5.38$, it provides useful constraints on $\Omega^{(\rm H I)}_{\rm IGM}$, the fraction of baryons found in the forest. To this end, we use three methods of calculating that quantity: the methods of \citet{pen00,pen04}, \citet{sch01}, and \citet{shu12}.

The method of \citet{pen00} assumes that Ly$\alpha$ forest absorbers are singular isothermal spheres and further assumes an impact parameter, $p$, scaled here as $p=(100 {\rm~ kpc})p_{100}$. As a result, the assumptions of this method are only valid for \ion{H}{1} absorbers at fairly low column density ($\log N\lesssim 14.5$). The method adopts an approximate, frequency-integrated \ion{H}{1} photoionization rate given by,
\begin{eqnarray}
\Gamma_{\rm HI} \approx (\unit{2.49\times 10^{-14}}{\reciprocal\second})J_{-23}\left (  \frac{4.8}{\alpha_s+3} \right ),
\end{eqnarray}
where $J_{-23}=1$ is the ionizing radiation field at 1 Rydberg in units of $10^{-23}\rm ~ergs ~cm^{-2} ~s^{-1}~ Hz~ sr^{-1}$ and $\alpha_s$ is the spectral index of the radiation field, set to $\alpha_s=1.8$. Thus, scaling the \ion{H}{1} column density as $N_{\rm HI}=(10^{14}~ {\rm cm^{-2}})N_{14}$, we can define
\begin{eqnarray}
\Omega^{(\rm H I)}_{\rm IGM} &=& \frac{(1.59\times 10^9 ~M_{\astrosun})H_0 \left ( { p_{100}^5 J_{-23} \left [ 4.8/(\alpha_s +3) \right ] }\right )^{1/2} }{c\pi p^2 \rho_{\rm cr}} \nonumber\\
&&\times~\int^{z_{\rm max}}_{0}\int^{N_{\rm max}}_{N_{\rm min}}  \frac{d\mathcal{N}( N_{\rm HI})}{dz}N_{14}^{1/2}d( N_{\rm HI})dz \nonumber \\
&=&8.73 \times 10^{-5}~h^{-1}_{70}\left (  J_{-23}~ p_{100} \frac{4.8}{\alpha_s+3}\right )^{1/2} \nonumber \\
&\times&~\sum_{j=0}^{z_{\rm max}}~\sum_{i=\log N_{\rm min}}^{\log N_{\rm max}}\left [ \frac{d\mathcal{N}(\log N_{\rm HI})}{dz}\right ]_{i,j} N_{14}^{1/2}\frac{\Delta z_j}{z_{\rm max}}.
\end{eqnarray}

Our preferred method for low column density absorbers, however, is the revised method of \citet{shu12}. Although this method retains the assumptions of \citet{pen00} regarding the scale and geometry of the IGM clouds, it also includes redshift-dependent corrections for the evolution in the space density of absorbers, $\phi (z) \propto (1+z)^3$, as well as the hydrogen photoionization rate. \citet{haa11} found a rapid increase in the metagalactic ionizing background radiation from $z=0$ to $z=0.7$, fitted by \citet{shu12} to the form,
\begin{eqnarray}
\Gamma_{\rm HI}&=(2.28\times 10^{-14} ~{\rm s^{-1}})(1+z)^{4.4}.
\end{eqnarray}
Thus, scaling the electron temperature as $T=(10^{4.3}~ {\rm K})T_{4.3}$, they obtain,
\begin{eqnarray}
\Omega^{(\rm H I)}_{\rm IGM} &=& (9.0 \times 10^{-5})~\frac{h^{-1}_{70}p_{100}^{1/2}T_{4.3}^{0.363}(1+z)^{0.2}}{\left [ \Omega_m(1+z)^3 +\Omega_{\Lambda}  \right ]^{1/2}}\nonumber \\
&\times&~\int^{z_{\rm max}}_{0}\int^{N_{\rm max}}_{N_{\rm min}}\frac{d\mathcal{N}(\log N_{\rm HI})}{dz}N_{14}^{1/2}d(\log N_{\rm HI})dz  \nonumber \\
&=&(9.0 \times 10^{-5})h^{-1}_{70}p_{100}^{1/2}T_{4.3}^{0.363} \nonumber \\
&\times&~\sum_{j=0}^{z_{\rm max}}~\sum_{i=\log N_{\rm min}}^{\log N_{\rm max}}\left ( \frac{(1+z_j)^{0.2}}{\left [ \Omega_m(1+z_j)^3 +\Omega_{\Lambda}  \right ]^{1/2}} \right. \nonumber \\
&&~~~~\times~\left.\left [ \frac{d\mathcal{N}(\log N_{\rm HI})}{dz}\right ] _{i,j}N_{14}^{1/2}\frac{\Delta z_j}{z_{\rm max}}\right ) .
\end{eqnarray}

Finally, we apply the method described by \citet{sch01}. This formalism assumes that the IGM clouds are gravitationally bound and that the observed column densities are characteristic over the local Jeans length. We expect that these assumptions better characterize high column density absorbers ($\log N_{\rm HI}\gtrsim 14.5$) than do the other methods. Hence, scaling the temperature as $T=(10^{4}~ {\rm K})T_{4}$, they define,
\begin{eqnarray}
\Omega^{(\rm H I)}_{\rm IGM} &=& (1.46 \times 10^{-4})~h^{-1}_{70}\Gamma_{-12}^{1/3}T_4^{0.59} \nonumber\\
&\times&~\int^{z_{\rm max}}_{0}\int^{N_{\rm max}}_{N_{\rm min}}\frac{d\mathcal{N}( N_{\rm HI})}{dz}N_{14}^{1/3}d( N_{\rm HI})dz  \nonumber \\
&=&(1.46 \times 10^{-4})~h^{-1}_{70}\Gamma_{-12}^{1/3}T_4^{0.59} \nonumber \\
&\times&~\sum_{j=0}^{z_{\rm max}}~\sum_{i=\log N_{\rm min}}^{\log N_{\rm max}}\left ( \left [ \frac{d\mathcal{N}(\log N_{\rm HI})}{dz}\right ]_{i,j} \right. \nonumber \\
&&~~~~\times~\left.  N_{14}^{1/3}\frac{\Delta z_j}{z_{\rm max}}\right ),
\end{eqnarray}
where we set $\Gamma_{\rm HI}=0.03\times 10^{-12}~ \rm s^{-1}$ at $z\sim 0$ \citep{shu99, wey01}.

For all methods, we assume a temperature of $T=2\times10^4~\rm K$ and sum over bins in column density and redshift. We set $p_{100}=1$ and normalize to $N_{14}=1$ throughout.  For our $\Delta z$ pathlength in Ly$\alpha$, cosmic variance is expected to contribute insignificantly to the errors, which instead are dominated by Poisson error  \citep{pen04}. We therefore calculate the error in $\Omega^{(\rm H I)}_{\rm IGM}$ as the one-sided Poisson error \citep{geh86} in $d\mathcal{N}(d\log N,  dz)$ bins summed in quadrature.

\input{sec243}

\section{Results and Discussion}

\subsection{Properties of the Catalog}
\subsubsection{Comparison to Previous \ion{O}{6} Catalogs}
\label{sec:catdiff}
\begin{figure*}
\plotone{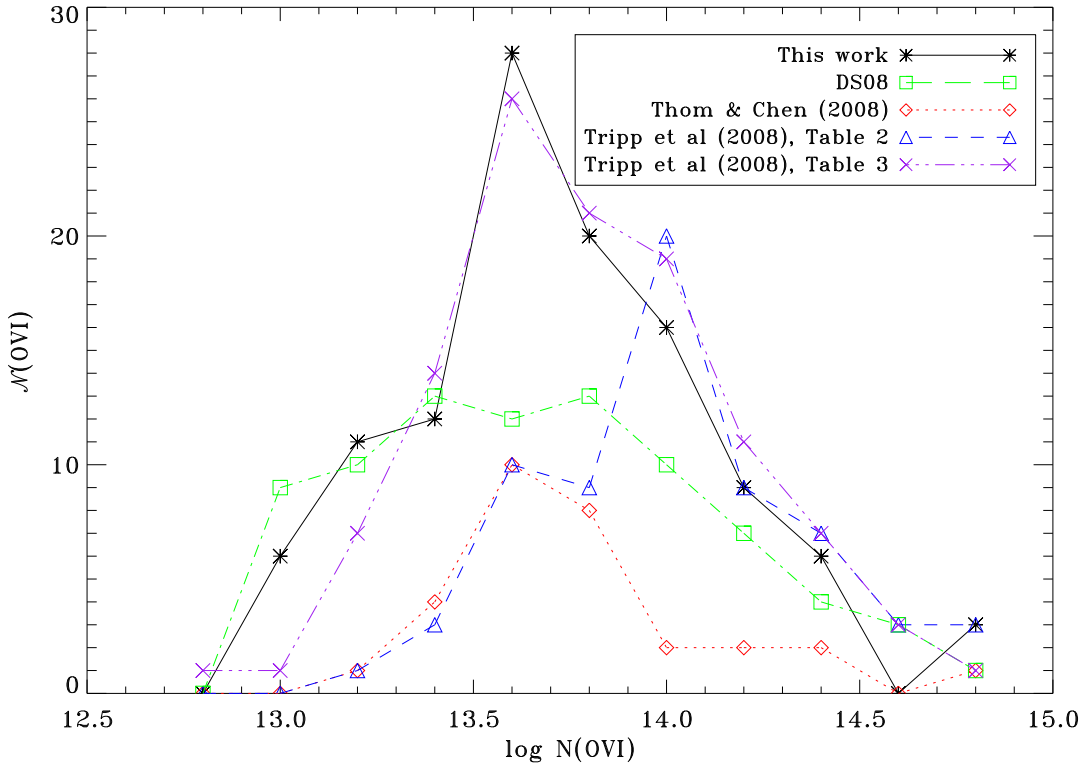}
\caption{Histograms comparing \ion{O}{6} detections as reported in this work, \citetalias{dan08}, \citet{tri08}, and \citet{tho08}. Bins in $\log N_{\rm OVI}$ are 0.2 dex in width, centered on the plotted points. The two separate sets of points for \citet{tri08} represent their ``system'' (Table 2) and ``component'' (Table 3) measurements.  Our catalog and component results of \citet{tri08} are strongly peaked at $\log N_{\rm OVI}\approx 13.6$. \label{fig:ovicats}}
\end{figure*}
As previously discussed, a number of studies have been undertaken to study the WHIM as traced by FUV \ion{O}{6} absorption, most notably \citetalias{dan08}, \citet{tri08}, and \citet{tho08}. Each of thse studies broadly used the same STIS data sets, albeit with independent reduction processes, some minor variation in sight-line selection, and with various degrees of assistance from \textit{FUSE} data. Their line selection and measurement methodologies, however, vary considerably, and produce significantly different catalogs. Figure~\ref{fig:ovicats} presents column-density histograms of the \ion{O}{6} detections reported by each of those catalogs, as well as the catalog presented here. As these variations can lead to different inferences and derived quantities, they must be kept in mind when interpreting this work or comparing studies of the WHIM. Though this discussion focuses on \ion{O}{6}, it is analogous to the distributions of other ion species.

The differences in the catalogs can be attributed primarily to the line selection criteria as well as approach to splitting detected features into subcomponents. In terms of detection criteria, \citetalias{dan08} took an inclusive approach indexed to Ly$\alpha$, in which any feature detected at 4 $\sigma$ or greater at close to the appropriate wavelength was included. \citetalias{dan08}'s approach to component splitting, however, was conservative: most blended lines were reported as a single absorption feature and most clearly split metal lines were reported together if the associated \ion{H}{1} absorption was of ambiguous structure. Together, these approaches lead to more detections at low column densities (some of which may be erroneous identifications), fewer detections at intermediate column densities, and a flatter histogram when compared to other catalogs.  In addition, a few \ion{O}{6} absorbers with little or no \ion{H}{1} absorption at the same redshift were missed in that study. In contrast, \citet{tho08} took a very conservative approach to inclusion, requiring that both lines of the \ion{O}{6} doublet be detected blindly without reference to \ion{H}{1}. This leads to fewer detections overall, especially at low column densities. Subcomponents are split more liberally than in \citetalias{dan08}. Finally, \citet{tri08} took an approach that combined the two aforementioned methods. The \ion{O}{6} doublets were identified blindly before a second search pass looked for \ion{O}{6} at redshifts suggested by strong absorption from other ions. The issue of subcomponents was avoided by presenting two sets of measurements: one that considered absorption ``systems'' with no splitting and one that split the systems into all discernable components. As a result, the histogram for their ``component'' detections is strongly peaked at intermediate column densities, while their systemic measurements lead to a distribution shifted to high column densities and reduced in total number of detections.

As described in Section~\ref{sec:construct}, our catalog is intermediate in its approach to component splitting compared to the aformentioned studies. It includes all reasonable detections included in previous catalogs that are not contradicted by new COS data. As a result, it is the largest catalog in terms of number of \ion{H}{1} and \ion{O}{6} detections. The \ion{O}{6} distribution is strongly peaked at $\log N\approx 13.2$, and it contains an intermediate number of detections at lower column densities compared to \citetalias{dan08} and \citet{tri08}. Compared to \citetalias{dan08}, it corrects a number of misidentifications at low column densities and augments the number of detections in light of new data and work by other research groups. Our catalog has 96 more \ion{H}{1} absorbers and 28 more \ion{O}{6} absorbers than \citetalias{dan08}. The increase can be attributed to the inclusion of new absorbers and sight lines as well as the more liberal splitting of absorbers into separate features. The corrected misidentifications were primarily weak metal lines mistaken for other species (e.g., \ion{Si}{3} mistaken for Ly$\alpha$) or uncommonly seen metals associated with high column density \ion{H}{1} absorbers mistaken for \ion{H}{1}.  As a result, this catalog supercedes that of \citetalias{dan08} for most purposes.

\subsubsection{Contamination in reported absorbers in STIS data}
\begin{figure}
\epsscale{1.2}
\plotone{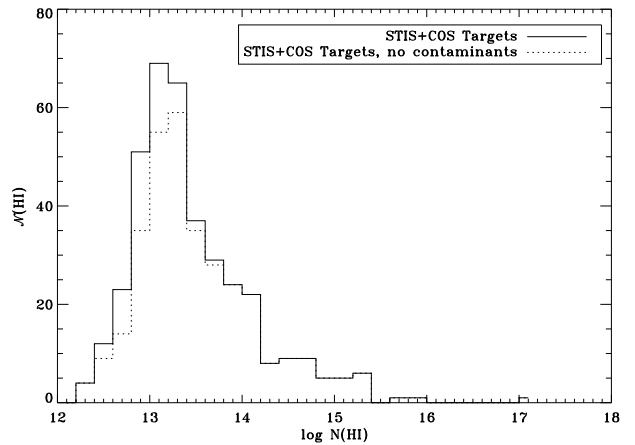}
\caption{Histogram comparing detected \ion{H}{1} absorbers for sight lines with both STIS and COS data. The solid line represents all detections, while the dotted line represents detections seen in both data sets.\label{fig:cats}}
\end{figure}
Our catalog contains 381 \ion{H}{1} detections in the 20 sightlines for which both COS and STIS and/or FUSE data are present. Of these detections, only 330 are detected by COS as well as STIS, as depicted in Figure~\ref{fig:cats}. In most cases, because the COS data have much higher S/N than the STIS data, this discrepancy suggests that some absorbers previously reported in the literature are in fact noise features rather than IGM absorption lines. For column densities $N_{\rm HI}\lesssim10^{13.2} \rm ~cm^{-2}$, between 20\% and 40\% of STIS detections are not supported by the COS data. In the column density range $10^{13.2}-10^{13.6}\rm ~cm^{-2}$, this improves to around 5\% to 10\% contamination, while higher column densities are nearly entirely free from contamination. These results suggest that censuses of the low-redshift Ly$\alpha$ forest at column densities below $10^{13.5} \rm ~cm^{-2}$ may suffer significant contamination in the form of line misidentification that is not accounted for in standard completeness corrections. Such contamination underscores the need for high-S/N observations of the IGM sensitive to Ly$\alpha$ equivalent widths $W_{\lambda}\approx \unit{5}{\milli\angstrom}$, corresponding to $\log N_{\rm HI}\approx 12.0$.

\subsection{Properties of the Absorbers}
Table~\ref{tab:detect} summarizes our detection statistics for several ion species, listed in Column (1). The ions in the bottom half of the table have poor statistics, and should be viewed with skepticism. Columns (2) and (3) report the total number of 4 $\sigma$ detections in STIS data and the number of detections with false positives indicated by COS data removed, respectively. The latter number is the set of detections used in all statistics and derived quantities. Column (4) lists the range of redshifts that the survey covers for a given ion, while Column (5) lists the maximum redshift pathlength sampled at any sensitivity for the ion. Columns (6) through (8) report the absorber frequency per unit redshift pathlength integrated down to three equivalent widths. The survey is greater than 50\%  complete at the \unit{30}{\milli\angstrom} scale, but at lower equivalent widths the completeness corrections take on a larger role. Finally, Column (9) provides a fit, in terms of the index $\beta$, to the distribution of absorbers as a function of column density per unit redshift,
\begin{eqnarray}
\frac{d^2\mathcal{N}(>N)}{dz~d\log N}\propto N^{-\beta},
\end{eqnarray}
where we have used an error-weighted fit to the cumulative distributed histogram of $d\mathcal{N}/dz$. This distributed approach may systematically flatten the slope, $\beta$, by $0.05-0.1$ dex relative to simple histograms \citepalias{dan08}.

In Table~\ref{tab:omega}, we summarize our results for the baryon fraction for metal ions, $\Omega_{\rm IGM}^{\rm (ion)}$ and $\Omega_{\rm ion}$, listed for equivalent widths integrated down to \unit{10}{\milli\angstrom} and \unit{30}{\milli\angstrom}. $\Omega_{\rm IGM}^{\rm (ion)}$ is scaled by metallicity and peak ionization fraction as described above, and it is reported relative to the total baryon fraction,  $\Omega_b=0.0455\pm0.0028$ \citep{kom11}.
\subsubsection{Individual Species}

\paragraph{\ion{H}{1}.}
\begin{figure*}
\plotone{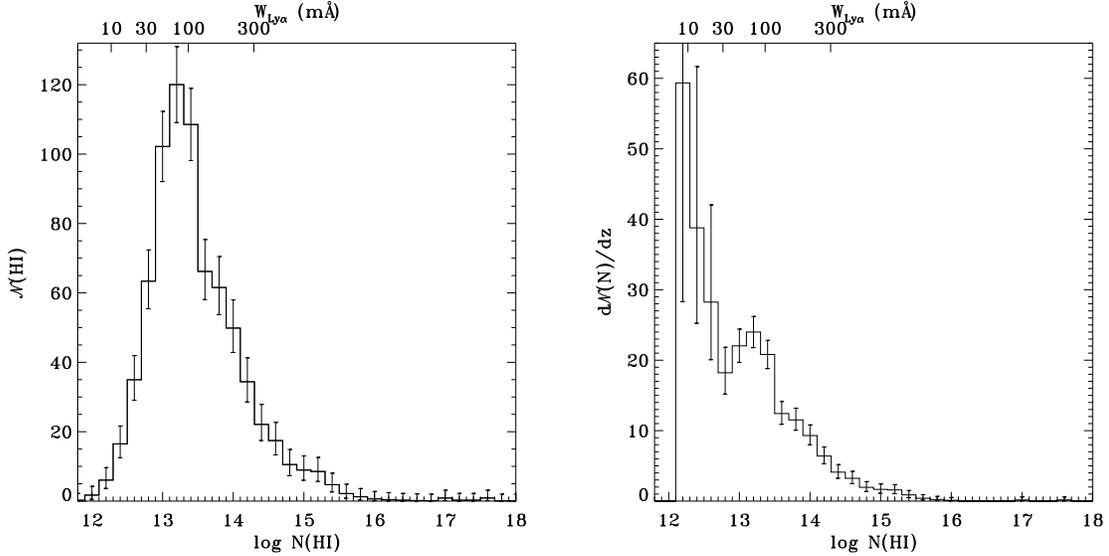}
\caption{Distributed histograms of \ion{H}{1} detection statistics and $\mathcal{N}({\rm H\,I})\equiv d\mathcal{N}/dz$. Left panel shows the number of absorbers per 0.2 dex bin, uncorrected for completeness. Right panel shows the completeness-corrected differential $d\mathcal{N}/dz$ vs. $\log N_{\rm HI}$. Error bars in both panels include one-sided Poisson errors, while the right panel errors also include contributions from $\Delta z$ uncertainty.\label{fig:hidist}}
\end{figure*}
Our results for the column-density distribution of \ion{H}{1} absorbers are presented in Figure~\ref{fig:hidist}. The histogram and $d\mathcal{N}/dz$ plots both exhibit a dip around $\log N_{\rm HI}\approx 13.5$, which was also seen in \citetalias{dan08}. The most likely explanation for this feature is that it is an artifact of the measurements arising from the detectability of higher-order Lyman lines.  There is therefore more frequent use of COG measurements at column densities beyond this point. As discussed in \citet{dan06}, measurements of absorbers that exhibit only Ly$\alpha$ may be biased toward lower column densities. This effect may cause some absorbers to scatter into lower bins, creating the dip in the distribution near the transition between Ly$\alpha$-only measurements and COG measurements. We expect that this effect has little impact on our absorber statistics, since it does not affect the total number of detections. A second dip is seen in the $d\mathcal{N}/dz$ plot at $\log N_{\rm HI} \approx 12.8$. It is not significant at the $95\%$ confidence level, and the incompleteness correction dominates in this regime. 

The entire distribution yields a power-law index, $\beta=1.68\pm0.03$, in agreement with  \citet{pen04}, who found $\beta=1.65\pm0.07$; \citet{leh07}, who found $\beta=1.76\pm0.06$; and \citetalias{dan08}, who found $\beta=1.73\pm0.04$. Limiting the fit to bins that are greater than 50\% complete, however, yields a slightly steeper value, $\beta=1.74\pm0.04$, which may indicate systematic effects from our completeness corrections. The completeness-corrected line frequency integrated down to \unit{30}{\milli\angstrom} ($\log N_{\rm HI}\approx 12.74$) is $d\mathcal{N}/dz=144^{+7}_{-6}$, in good agreement with \citet{pen04}. \citetalias{dan08} found a lower value of $d\mathcal{N}/dz=129^{+6}_{-5}$, where the discrepancy likely arises from the methodological differences described in Section~\ref{sec:catdiff}.

\begin{figure*}
\plotone{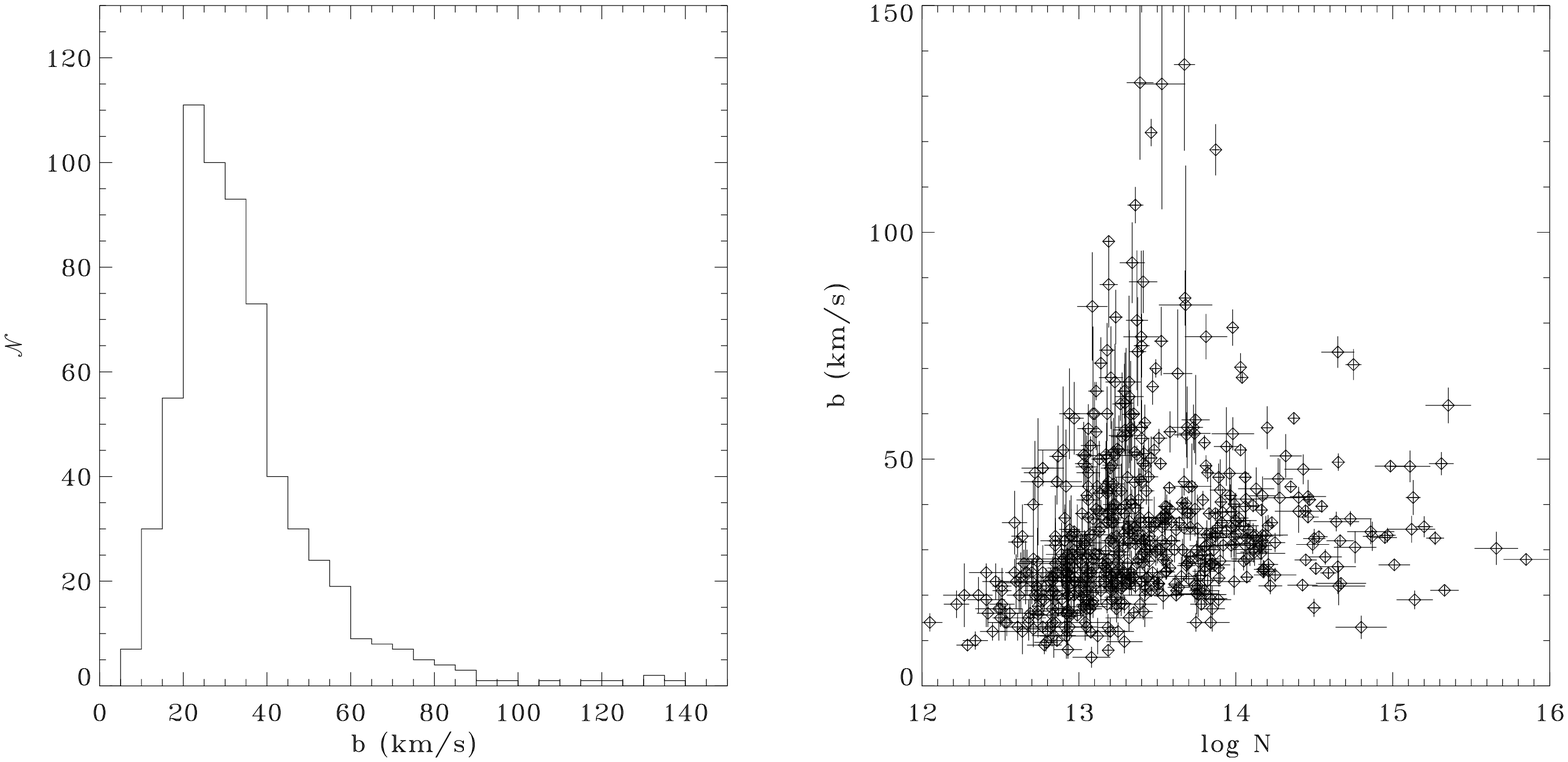}
\caption{Distribution of Doppler $b$-parameters of 626 of 746 \ion{H}{1} absorbers, limited to absorbers with errors in $b$ and $N_{\rm HI}$ less than 50\%. The median and sample standard deviation of the distribution are \unit{30}{\kilo\meter\usk\reciprocal\second} and \unit{17}{\kilo\meter\usk\reciprocal\second}, respectively. Because some absorbers may actually be multiple blended components, this distribution provides upper limits on $b$-values.\label{fig:hibdist}}
\end{figure*}
The distribution of the Doppler parameter, $b$, for \ion{H}{1} provides constraints on the thermal state of the IGM. The distribution of measured $b$-values, limited to absorbers with errors in $b$ and $N_{\rm HI}$ less than 50\%, is shown in Figure~\ref{fig:hibdist}, with a range from $b=\unit{6-137}{\kilo\meter\usk\reciprocal\second}$ with a mean, median, and standard deviation of 34, 30, and \unit{17}{\kilo\meter\usk\reciprocal\second}, respectively. Values of $b=(\unit{30}{\kilo\meter~\reciprocal\second})b_{30}$ correspond to a thermal temperature of $T\approx (\unit{54,500}{\kelvin})b^2_{30}$, but this number can be taken only as an upper limit because turbulence, bulk flows, and undetected subcomponents will inflate the measurements. The broad Ly$\alpha$ absorbers (BLAs) predicted by simulations and often associated with the WHIM \citep[e.g.,][]{ric06} are difficult to characterize in the noisy STIS echelle data. Roughly 25\% of our \ion{H}{1} catalog has a measured $b>\unit{40}{\kilo\meter\usk\reciprocal\second}$, but the errors are large in this regime. In addition to the large statistical errors, systematic effects such as unresolved subcomponents, insufficiently characterized AGN continua, and instrumental artifacts likely add further uncertainty to the upper end of the distribution of $b$-values. The treatment of BLAs is dealt with more carefully in \citet{leh07} and \citet{dan10}, with the latter utilizing much higher S/N data from COS to further constrain their properties; we refer the interested readers to those works.

\paragraph{\ion{O}{6}.}
\label{sec:o6}
\begin{figure*}
\plotone{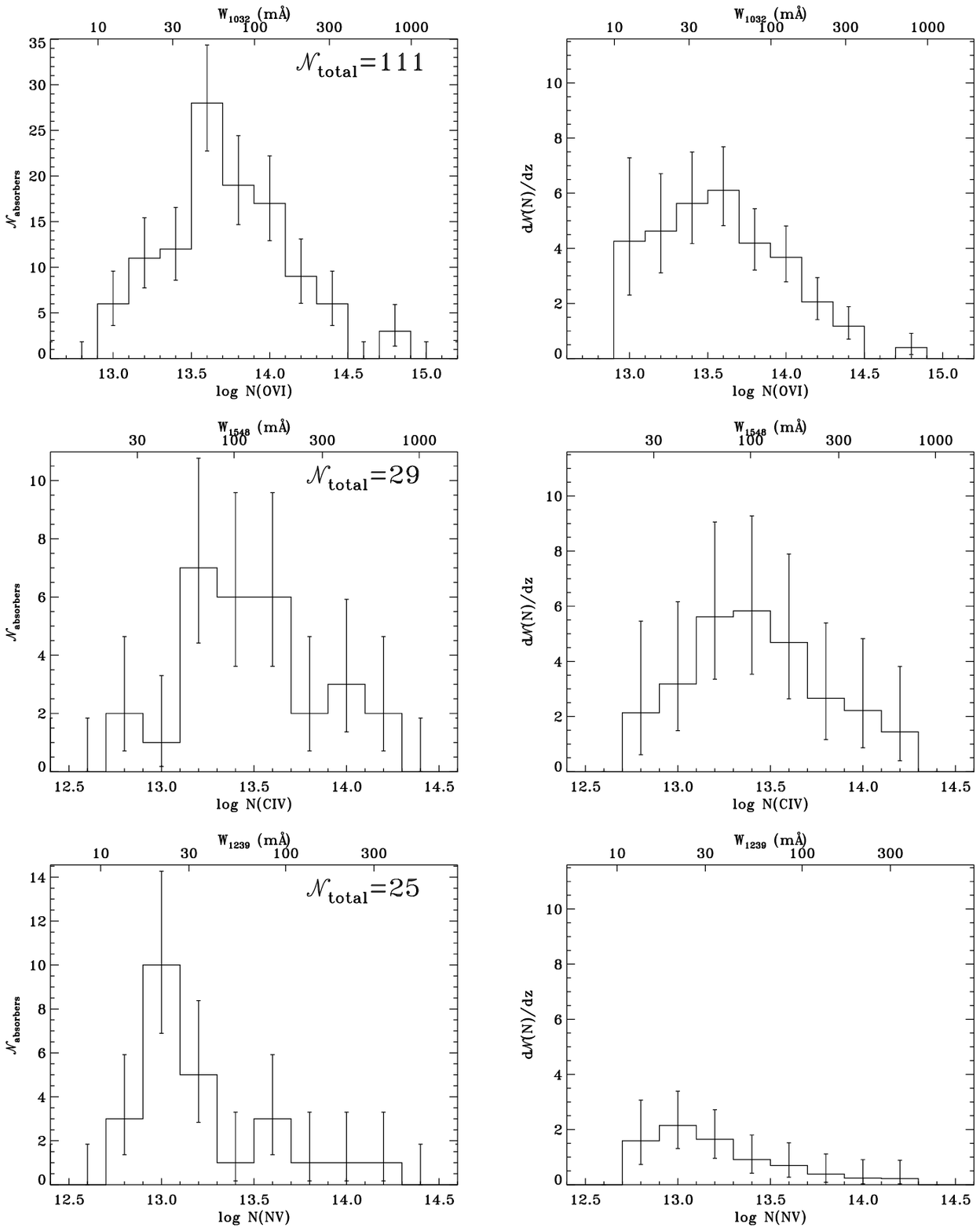}
\caption{Simple histograms of detection statistics and distributed histograms of $d\mathcal{N}/dz$ vs. $\log N$ for \ion{O}{6} (top), \ion{C}{4} (middle), and \ion{N}{5} (bottom).\label{fig:highions}}
\end{figure*}
We find a power-law index for the entire \ion{O}{6} distribution (Figure~\ref{fig:highions}, top panels) of $\beta=2.08\pm0.12$, in good agreement with our group's previous work \citep{dan08, dan05}. Interestingly, we see a turnover in the $d\mathcal{N}/dz$ column density distribution at $\log N_{\rm OVI}\approx 13.5$. This effect was also seen by \citet{dan05}, albeit with just five absorbers at such low column densities. On the other hand, \citetalias{dan08} found no evidence of a turnover, with 26 absorbers in the bins at $\log N_{\rm OVI}< 13.5$. We report 29 absorbers in our three lowest bins ($12.9<\log N_{\rm OVI} <13.5$) over a larger pathlength in $z$. The turnover seems to arise from the combined effect of a stronger peak in the distribution of absorbers at $\log N_{\rm OVI} \approx 13.6$ coupled with an increase of only 3 absorbers with $\log N_{\rm OVI}< 13.5$ compared to \citetalias{dan08}. Our correction of several absorbers from the \citetalias{dan08} catalog that were misidentified as \ion{O}{6} explains the lack of a large increase in the number of absorbers in these bins despite our larger pathlength. Despite its statistical significance, the turnover should be treated with caution. The statistical errors on individual measurements at these column densities are quite large, and the method of catalog construction may introduce systematic biases at low column densities owing to the predominant reliance on the catalog of \citetalias{dan08} in that regime. It is also worth noting that our completeness corrections exceed 50\% in these bins. A new COS survey at higher S/N will be required to conclusively address the issue of this turnover. If the turnover is real, it may be more realistic to apply the power-law formalism only to data with $\log N_{\rm OVI}> 13.5$; this approach yields a much steeper slope, $\beta=2.50\pm0.26$.

\begin{figure}
\epsscale{1.2}
\plotone{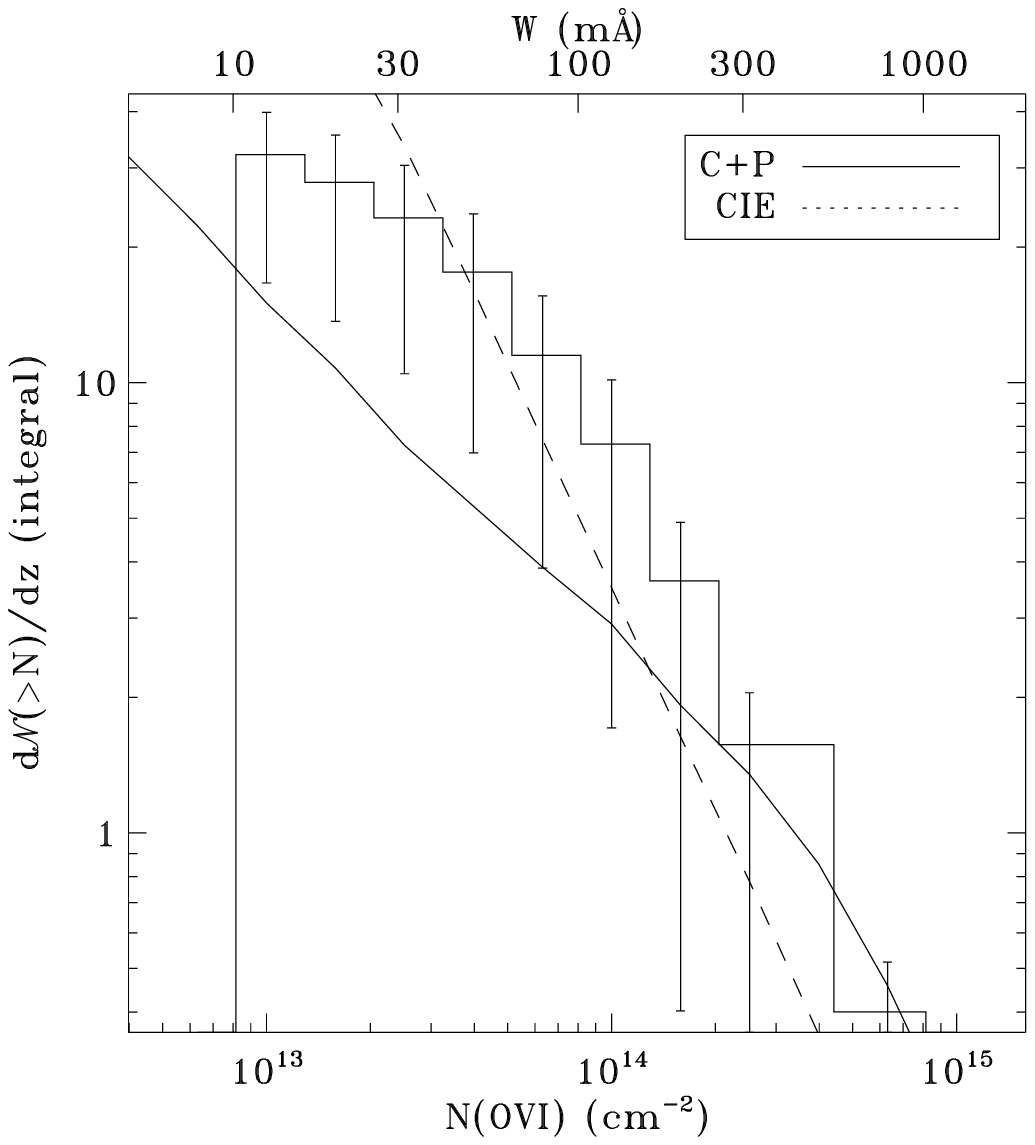}
\caption{Cumulative $d\mathcal{N}(>N)/dz$ for \ion{O}{6} absorbers compared to simulation results of run 50\_1024\_2 from \citet{smi11} featuring collisional and photoionization (C+P - solid line) and CIE cooling (dashed line).\label{fig:o6smith}}
\end{figure}
Our results for the frequency of \ion{O}{6} absorbers compare favorably to simulations. \citet{smi11} investigated the distribution using the N-body hydrodynamic code Enzo in a $50h^{-1}~\rm Mpc$ comoving box with $1024^3$ grid cells (run 50\_1024\_2 in that work). Cooling in CIE and with both collisional and photoionization (C+P) were investigated. We plot their results for the cumulative distribution of $d\mathcal{N}(>N_{\rm OVI})/dz$ against our empirical distribution in Figure~\ref{fig:o6smith}. Though CIE adequately explains the distribution at high column densities, it over-predicts the distribution at lower column densities, suggesting an increasing role of photoionization at these scales. These smaller clouds may reside in areas of the IGM that are characterized by overall lower baryonic densities, making them more sensitive to the effects of photoionization. Alternately, perhaps they reside in the circumgalactic medium, where higher UV radiation fields may increase the role of photoionization.

\begin{figure*}
\plotone{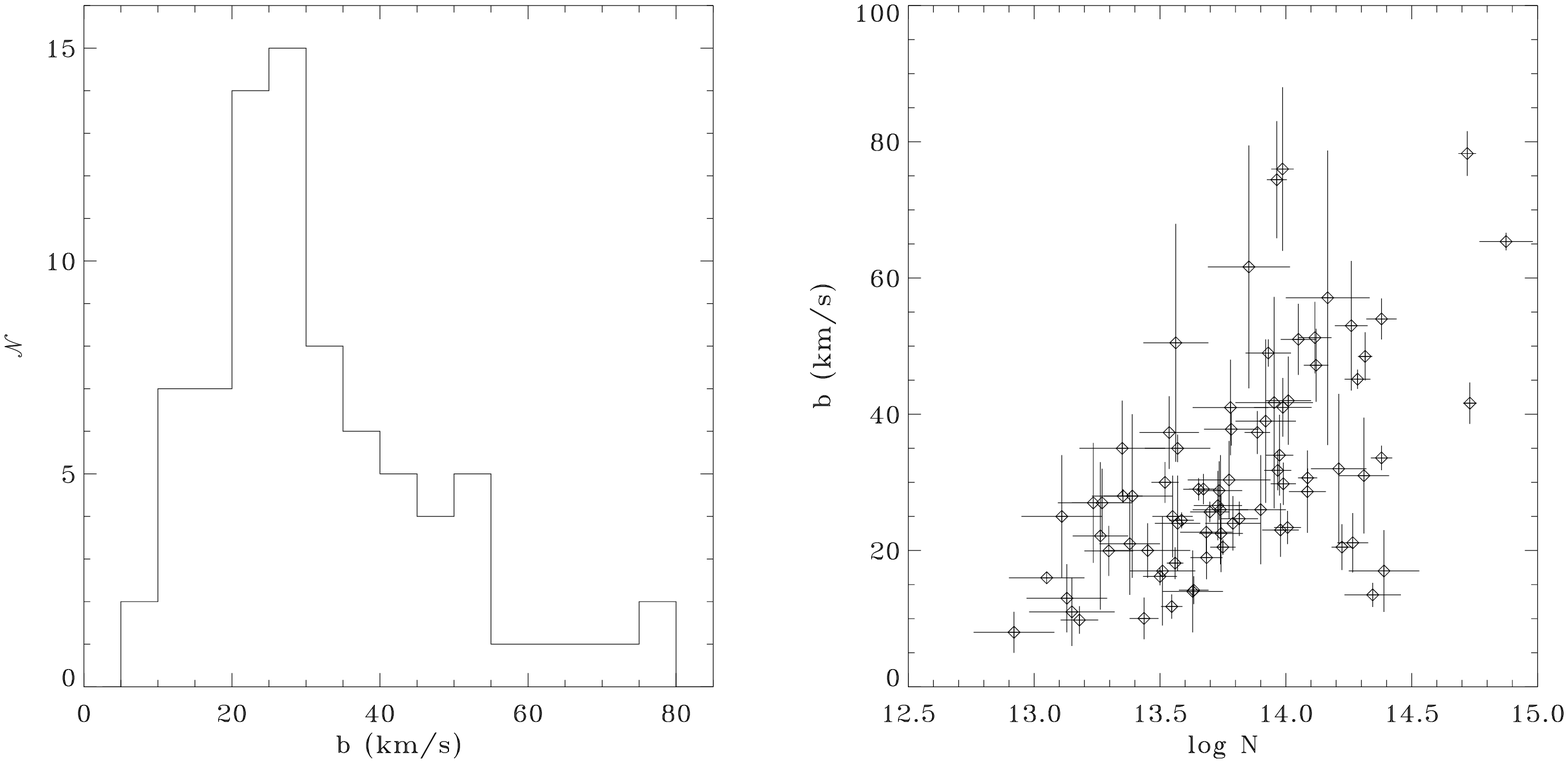}
\caption{Distribution of Doppler $b$-parameters of 79 of 111 \ion{O}{6} absorbers, limited to absorbers with errors in $b$ and $N$ less than 50\%. The median and sample standard deviation of the distribution are \unit{28}{\kilo\meter\usk\reciprocal\second} and \unit{16}{\kilo\meter\usk\reciprocal\second}, respectively. Because some absorbers may actually be multiple blended components, this distribution provides upper limits on $b$-values.\label{fig:ovibdist}}
\end{figure*}
In Figure~\ref{fig:ovibdist}, we present the \ion{O}{6} Doppler $b$-parameter distribution, again limited to absorbers with errors in $b$ and $N_{\rm OVI}$ less than 50\%. Nearly all measurements in this distribution were performed using the the Voigt profile routine described previously, which accounts for instrumental resolution. The distribution spans a range of $b=\unit{6-78}{\kilo\meter\usk\reciprocal\second}$ with a mean, median, and standard deviation of 32, 28, and \unit{16}{\kilo\meter\usk\reciprocal\second}, respectively, in excellent agreement with \citetalias{dan08}. Qualitatively, the distribution is similar to Figure~13 of \citet{tri08}, although they find a higher density of absorbers with $b<\unit{10}{\kilo\meter\usk\reciprocal\second}$. We suspect that this discrepancy arises from their more aggressive splitting of systems into subcomponents, which may artificially lower their distribution of $b$-values.  Eleven absorbers (14\%) in our catalog have $b$-values corresponding to temperatures at or below $\unit{10^{5.45}}{\kelvin}$, the peak in CIE. Only two (3\%) correspond to temperatures less than $10^5$~K as expected from photoionization.

\begin{figure}
\epsscale{1.18}
\plotone{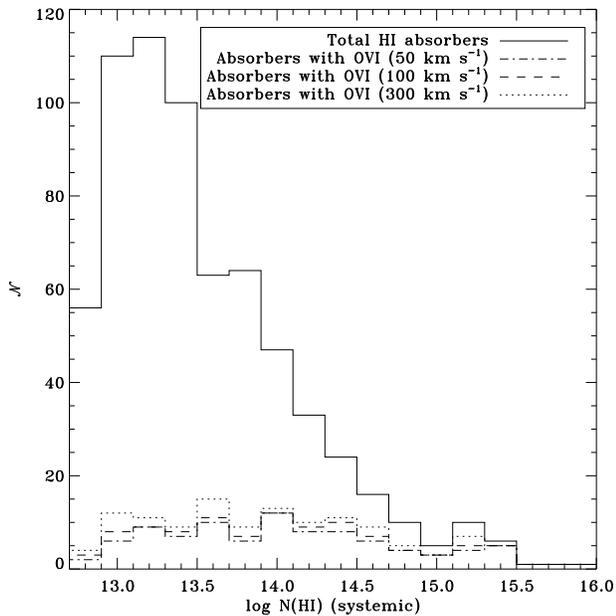}
\caption{Histograms of all ``systemic" \ion{H}{1} absorbers (see Section~\ref{sec:o6} for details), in which subcomponents within \unit{50}{\kilo\meter\usk\reciprocal\second} of each other have been combined (solid line), and systems with \ion{O}{6} absorbers within \unit{50}{\kilo\meter\usk\reciprocal\second} (dotted line), \unit{100}{\kilo\meter\usk\reciprocal\second} (dashed line), and \unit{200}{\kilo\meter\usk\reciprocal\second} (dash dotted line) of the systemic \ion{H}{1} redshift. \label{fig:ovihi}}
\end{figure}
To investigate the correlation of \ion{O}{6} absorption with \ion{H}{1} absorption, we combined all components of each ion species with $|\Delta v| \le \unit{50}{\kilo\meter\usk\reciprocal\second}$, yielding ``systemic" column densities and velocity displacements. We then correlated the two set of absorbers, counting any \ion{O}{6} absorption with a redshift within several $\Delta v$ offsets of the redshift of the \ion{H}{1} as being associated with that \ion{H}{1} absorber. The resulting distributions are plotted as histograms as a function of column density in Figure~\ref{fig:ovihi}. The strongest \ion{H}{1} absorption systems nearly always have associated \ion{O}{6}. In all but a handful of cases, \ion{O}{6} has associated \ion{H}{1} absorption. We found four cases of \ion{O}{6} absorption without apparent associated \ion{H}{1} absorption within $|\Delta v|\le\unit{100}{\kilo\meter\usk\reciprocal\second}$ and only one case without \ion{H}{1} absorption within $|\Delta v|\le\unit{300}{\kilo\meter\usk\reciprocal\second}$. This kinematic association confirms that \ion{H}{1} and \ion{O}{6} often arise in spatially associated, multiphase systems, as predicted by simulations \citep{cen99,dav01,smi11} and observed \citep{dan08, tri08,sav11b}.

\paragraph{\ion{C}{4} and  \ion{N}{5}.}
The Li-like \ion{C}{4} and \ion{N}{5} ions have also been suggested as possible tracers of WHIM gas, although photoionization probably also plays a role for these species. We report 29 \ion{C}{4} absorbers, shown in Figure~\ref{fig:highions} (middle row), but our survey is sensitive to \ion{C}{4} over a much shorter total redshift pathlength ($\Delta z=2.52$) compared to other ions. The distribution yields a power-law index of $\beta =1.77\pm0.16$ and an absorber frequency  integrated down to \unit{30}{\milli\angstrom} ($\lambda 1548.2$; $\log N_{\rm CIV}\approx 12.87$) of $d\mathcal{N}/dz=12^{+4}_{-2}$. Our \ion{C}{4} catalog has changed little since \citetalias{dan08}, so, unsurprisingly, these results are in good agreement with the results from that study and its discussion of the ion. As in the \ion{O}{6} distribution, we see a turnover in the $d\mathcal{N}/dz$ column density distribution of \ion{C}{4} at $\log N_{\rm CIV}\approx 13.2$, but it is not significant at the 95\% confidence level. We invoke all of the same caveats in ascribing meaning to this feature as we did for the \ion{O}{6} results. 

Despite having more than double the redshift pathlength of \ion{C}{4}, the \ion{N}{5} ion has only 25 absorbers, shown in Figure~\ref{fig:highions} (bottom row), yielding a power-law index of $\beta =2.00\pm0.24$ and an absorber frequency integrated down to \unit{30}{\milli\angstrom} ($\lambda 1238.2$; $\log N_{\rm NV}\approx 13.15$) of $d\mathcal{N}/dz=10^{+5}_{-3}$. The low frequency of detections likely is attributable to the low abundance of nitrogen compared to oxygen or carbon. As a result, most detections are weak, and it will remain difficult to draw many conclusions from \ion{N}{5} absorption until a systematic survey in more sensitive COS data is undertaken. The slight turnover apparent in the $d\mathcal{N}/dz$ column density distribution is not statistically significant.

\paragraph{\ion{C}{3}, \ion{Si}{3}, \ion{Si}{4}, \ion{Fe}{3} and other ions.}
\begin{figure*}
\plotone{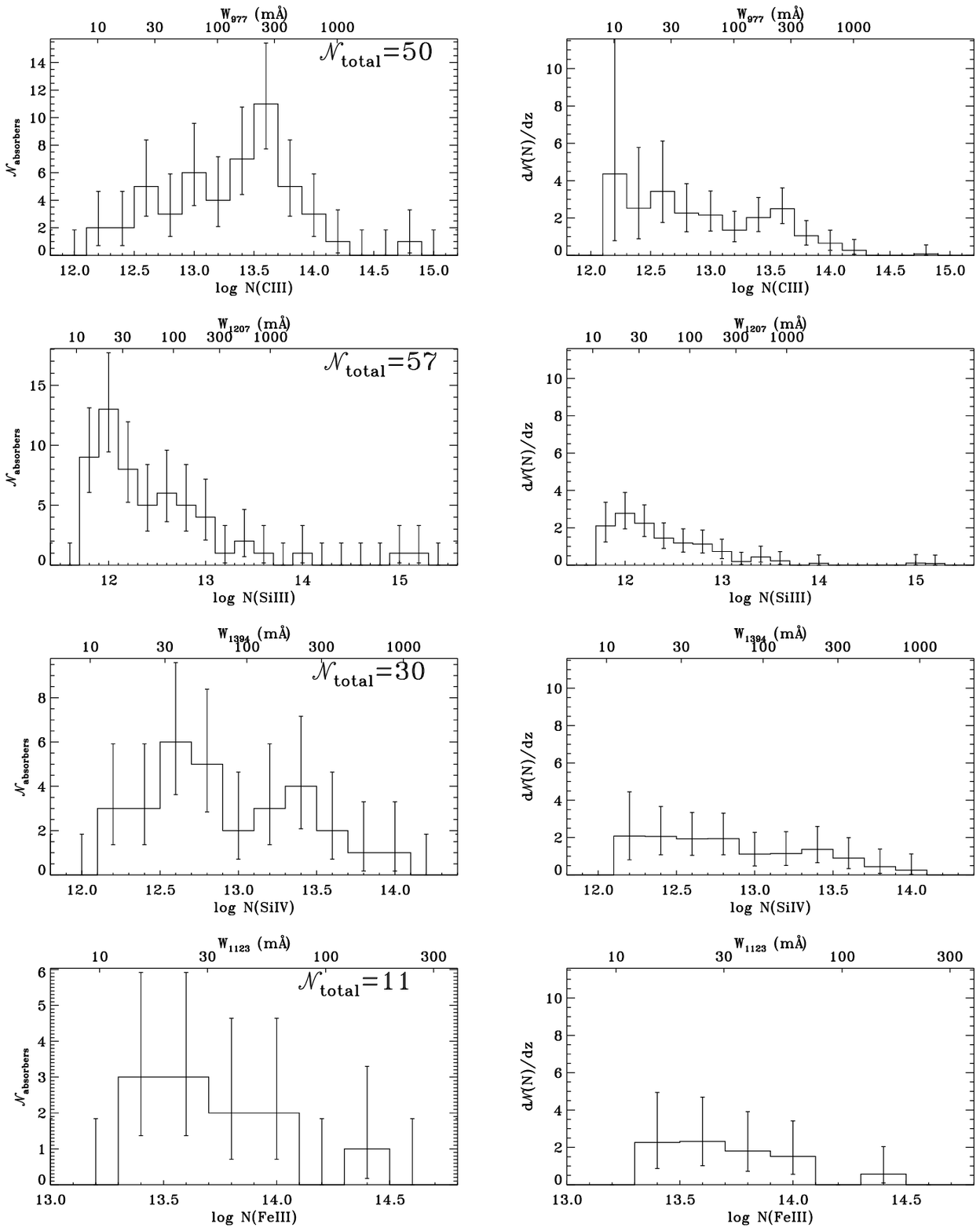}
\caption{Simple histograms of detection statistics and distributed histograms of $d\mathcal{N}/dz$ vs. $\log N$ for \ion{C}{3} (top), \ion{Si}{3} (second row), \ion{Si}{4} (third row), and \ion{N}{5} (bottom).\label{fig:lowions}}
\end{figure*}
\ion{C}{3} and \ion{Si}{3} are expected to arise in photoionized conditions, and they are well characterized by the high-resolution STIS data, owing to their relatively narrow line widths and strong absorption features. Our catalog's primary differences from that of \citetalias{dan08} arise from a more careful treatment of the narrow subcomponents seen in many of these absorbers.  We report 50 \ion{C}{3} absorbers, shown in Figure~\ref{fig:lowions} (top row), yielding a power-law index of $\beta =1.86\pm0.09$ and an absorber frequency integrated down to \unit{30}{\milli\angstrom} ($\log N_{\rm CIII}\approx 12.67$) of $d\mathcal{N}/dz=14.4^{+4.7}_{-2.2}$. We find 57 \ion{Si}{3} absorbers, shown in Figure~\ref{fig:lowions} (second row), yielding a power-law index of $\beta =1.74\pm0.10$ and an absorber frequency integrated down to \unit{30}{\milli\angstrom} ($\log N_{\rm SiIII}\approx 12.14$) of $d\mathcal{N}/dz=7.4^{+2.1}_{-1.2}$. We see no significant turnover in their $d\mathcal{N}/dz$ column density distributions.

We report 30 \ion{Si}{4} detections, shown in Figure~\ref{fig:lowions} (third row), which follow a distribution quite similar to that found in \citetalias{dan08}. They found a power-law index of $\beta=1.92\pm0.17$ that was somewhat steeper than that of the other low-ionization state species. Our catalog yields an index of $\beta=1.73\pm0.13$, bringing it more in line with the other ions. We find an absorber frequency integrated down to \unit{30}{\milli\angstrom} ($\lambda 1393.8$; $\log N_{\rm SiIV}\approx 12.53$) of $d\mathcal{N}/dz=9.2^{+4.5}_{-2.8}$.

\citetalias{dan08} made an ambitious attempt to characterize \ion{Fe}{3} absorption in the IGM, reporting 14 detections comprising a steep distibution with $\beta=2.2\pm0.4$. However, they expressed some caution about the validity of these results, owing to the high \ion{Fe}{3} abundance ratios. Our reanalysis of the data suggests that all \ion{Fe}{3} results should be treated with skepticism. Although we report 11 detections at the nominal 4 $\sigma$ level, nearly all detections for which COS data were available were revealed as misinterpreted noise features. It is thus likely that a significant number of those 11 absorbers may  vanish, too, if observed at a higher S/N with COS. In reporting our \ion{Fe}{3} statistics for this species in Table~\ref{tab:detect}, we tentatively note that they are poorly constrained.  Table~\ref{tab:detect} further reports our detections of several other low-ionization species. As with \ion{Fe}{3}, we mark them as poorly constrained owing to the paucity of detections. No statistics are presented for \ion{N}{2} ($\lambda 1084.0$, $\lambda 915.6$) or \ion{N}{3} ($\lambda 989.8$) because no systematic search was made for these species. The listed detections were primarily serendipitous corrections of past misidentifications in the literature, and they are primarily associated with high column density \ion{H}{1} absorption systems. 

\subsection{The Baryon Density of the IGM}
\label{sec:baryon}
Using the assumptions for metallicity and ionization fraction described above, the three best tracers of the WHIM yield estimates of the fractional baryon content that they trace as $\Omega_{\rm IGM}^{\rm (ion)}/\Omega_b\approx9\pm1\%$ (\ion{O}{6}), $6\pm 3\%$ (\ion{N}{5}), and $7\pm 1\%$ (\ion{C}{4}). \citetalias{dan08} offered several interpretations for variations in these values. Most obviously, the differences may arise due to dominant systematic errors arising from our assumed CIE ion fractions or metallicities, or they may arise simply due to line sensitivity variations in the data itself. That work also suggested that perhaps their values (9\%, 5\%, and 3\% for \ion{O}{6}, \ion{N}{5}, and \ion{C}{4}, respectively) represented a true gradient the WHIM baryon density as a function of temperature, because their derived densities follow the trend in peak CIE temperatures. This interpretation is not supported by our improved values, though we are unable to rule it out. 

The largest systematic uncertainties in these values arise from the metallicity and ionization fraction assumptions. In the above values, we assume a metallicity of 10\% solar and an ionization fraction corresponding to peak CIE value. Recent simulations \citep{shu12} suggest that the weighted mean product of these quantities may be closer to $f_{\rm OVI}(Z/Z_{\astrosun})=0.01$, half of our assumed value of 0.02. Using this number in our calculations effectively doubles the baryonic content traced by \ion{O}{6}, yielding $\Omega_{\rm IGM}^{\rm (ion)}/\Omega_b\approx 17.2\pm 1.4\%$. \citet{shu12} also report a column density dependent fit to this quantity: $f_{\rm OVI}(Z/Z_{\astrosun})=(0.015)[N_{\rm OVI}/10^{14}~{\rm cm^{-2}}]^{0.70}$. Evaluating that fit separately for each bin in our $\Omega$ calculations raises the baryonic content still further, to $\Omega_{\rm IGM}^{\rm (ion)}/\Omega_b\approx 18.9\pm 2.0\%$, but the scatter in the  $f_{\rm OVI}Z$  fit is roughly 0.35 dex. Simulations suggest, however, that the actual contribution of \ion{O}{6} to $\Omega_b$, as calculated by the direct summation of \ion{O}{6} within the simulation, may be as much as $10-40\%$ higher than that derived from absorption-line studies, as traced by the creation of synthetic spectra from the simulation \citep{smi11}. We can apply the analogous results from the same simulation, derived from run 50\_1024\_2 from \citet{smi11} (B.D. Smith, private communication), to our \ion{N}{5} and \ion{C}{4} results. Using the column density fits for these ions, we obtain baryon fractions of $10.1\pm6.0\%$ (\ion{N}{5}) and $22.3\pm3.7\%$ (\ion{C}{4}). The \ion{C}{4} results change most dramatically, possibly because it is likely to be more sensitive to photoionization effects. 

\input{omega_lya_allmethod}

The baryon content of the photoionized Ly$\alpha$ forest has been well studied elsewhere. \citet{pen04} accounted for roughly $29\%$ of the local baryon mass down to $\log N_{\rm HI} =12.5$, and \citet{leh07} found a similar estimate of around $30\%$. We present our own calculations based on the methods described in Section~\ref{sec:omega} in Table~\ref{tab:omega_lya}. The \citet{shu12} method is most appropriate at lower column densities, while the \citet{sch01} method, which assumes gravitationally bound clouds, is likely more appropriate at higher column densities. In estimating a total ratio $\Omega_{\rm Ly\alpha}/\Omega_b$, a cutoff between the two methods should be assumed. Taking a cutoff of $\log N_{\rm HI}=14.5$, corresponding to the transition in the slope of the column density distribution seen by \citet{pen04}, we find $\Omega_{\rm Ly\alpha}/\Omega_b=20.9\pm2.0\%$. If the cutoff is instead drawn at $\log N_{\rm HI}=16.5$, we find $\Omega_{\rm Ly\alpha}/\Omega_b=23.7\pm2.2\%$. Because this calculation is model dependent, systematic errors likely dominate over the statistical errors that we report.

\input{tablepaperomega}

The baryon fractions reported in Table~\ref{tab:omega} for the lower-ionization species likely trace a metal-enriched subset of the gas in the Ly$\alpha$ forest. As a result, they are probably photoionized, rendering assumptions regarding the ionization fraction problematic. Hence, the values of $\Omega_{\rm IGM}^{\rm (ion)}/\Omega_b$ for these ions are extremely uncertain beyond the statistical errors given in the table. Recalculating them with column density dependent fits from the \citet{smi11} simulation gives values 2 to 3 times higher, but this huge variation likely indicates imperfections in the combined ionization calculations in simulations to date.

\section{Conclusions and Summary}

We have presented a comprehensive, critically evaluated catalog of low-redshift IGM absorbers present in archival legacy STIS and FUSE data. New data from COS informed the evaluation, revealing past errors in the interpretation of absorption lines. We analyzed 44 sight-lines, yielding 746 \ion{H}{1} absorbers over a total Ly$\alpha$ redshift pathlength of $\Delta z=5.38$. In addition, we reported on 111 \ion{O}{6} absorbers, 29 \ion{C}{4} absorbers, and made numerous detections of other ion species (\ion{N}{5}, \ion{C}{3}, \ion{Si}{3}, \ion{Si}{4}).

Our \ion{O}{6} results found distributions comparable to those found in past work. We found a $d\mathcal{N}/dz$ turnover at low column densities similar to that seen by \citet{dan05} but absent in \citetalias{dan08}. Using a more sophisticated baryon accounting technique than previous work,  we estimated from our empirical distribution of absorbers that the WHIM contributes roughly 19\% of the total baryonic density, $\Omega_b$. We discussed several significant systematic uncertainties in deriving such a quantity, most notably corrections for the metallicity and ionization state of the IGM. We presented similar distributions and quanitities derived from \ion{N}{5} and \ion{C}{4} detections, of which we reported 25 and 29 absorbers, respectively. Similar statistics were reported for other ion species at lower ionization states, and we summarized these results in Table~\ref{tab:detect}. In particular, we found that past studies of \ion{Fe}{3} absorption in the low-redshift IGM suffer from significant contamination due to line misidentification.

Our catalog of 746 \ion{H}{1} absorbers follows a similar distribution to that of our group's past work in this field. \citetalias{dan08} used an incorrect cosmological factor to derive baryon densities from the detections; we corrected that error with our new catalog, accounting for roughly 24\% of $\Omega_b$ with Ly$\alpha$ forest detections. Taken together with inferred WHIM baryon densities, our catalog accounts for $\sim43\%$ of $\Omega_b$ in the low-redshift IGM. Unfortunately, the present catalog is unable to adequately probe broad Ly$\alpha$ absorbers, which are expected to trace significant baryonic content in the WHIM. Such absorbers require high S/N data, better supplied by future work with COS.

This paper has analyzed most of the IGM absorption-line systems at $z < 0.4$ observed by STIS.
With COS, one can take these studies much further through a
spectroscopic survey of the structure, metallicity, and physical conditions of the IGM and CGM.
Observing the relation of IGM absorption systems to galaxy distributions will be achieved through
uniformly processed, high-S/N spectra and an unbiased sample of IGM and CGM absorbers, including
the Ly$\alpha$ forest, metal-line systems, Lyman-limit systems, and partial-limit systems.
We have proposed a sensitive (S/N $\sim$ 30) HST Spectroscopic Legacy Survey of the low-redshift
IGM and CGM toward AGN selected for UV brightness and total redshift pathlength.   These absorbers
will be compared with deep galaxy surveys, thereby connecting galaxies to large-scale structure
and inferring the extent of feedback.   These projects will produce a complete legacy survey of gas and
($L > L_{\rm SMC}$) galaxies, located within 1-2 Mpc of the ($z = 0.1 - 0.4$) absorbers.  The high UV sensitivity of COS provides an
excellent tool for studying gas in multiple thermal phases and metallicities in the low-redshift IGM and CGM.
These studies are best done at low redshift ($z \leq 0.4$), allowing us to use correlations of the absorbers
with distributions of galaxies, particularly those of low mass and luminosity  ($L < L^*$) which contribute
significant radiative, thermal, and chemical feedback. While the catalog presented here represents an improvement to our census of IGM baryonic content, these future studies with COS will be able to address conclusively the distributions of absorbers at low column densities and broad line widths. Poorly constrained ions such as \ion{Fe}{3} and \ion{N}{5} may be probed with more reliability, and the increase in pathlength will dramatically improve the statistics of all ion species addressed here.

\acknowledgments

The authors would like to thank Thorsten Tepper-Garc\'{i}a for helpful comments regarding the errors in \citetalias{dan08} and Britton Smith for his contributions regarding simulation results for the metallicity and ionization state of the IGM. This work was supported by Space Telescope Science Institute Archive Legacy grant AR-11773.01, NSF grant AST07-07474, and STScI COS-support grant NNX08-AC14G. This work made use of observations made with the NASA/ESA \textit{Hubble Space Telescope}, obtained from the data archive at the STScI.  STScI is operated by the Association of Universities for Research in Astronomy, Inc., under NASA contract NAS5-26555.

{\it Facilities:} \facility{FUSE}, \facility{HST (COS, STIS)}.

\bibliographystyle{apj}
\bibliography{apj-jour,igmpaperbib}

\end{document}

%% file: ovi_sightlines.tex

\begin{deluxetable*}{lcclrrrrr}
\tabletypesize{\tiny}
\tablecaption{44 AGN Sightlines and data used in the catalog\label{tab:sightlines}}
\tablecolumns{8} 
\tablewidth{0pt} 
\tablehead{\colhead{Sight Line}   &
           \colhead{R.A.}         &
	   \colhead{Decl.}        & 
	   \colhead{$z_{\rm AGN}$}&
	   \colhead{STIS}         &
           \colhead{FUSE}         &
	   \colhead{COS/G130M}     &
	   \colhead{COS/G160M}     \\
	   \colhead{}             &        
	   \colhead{(J2000.0)}    &
	   \colhead{(J2000.0)}    &
	   \colhead{}             & 
	   \colhead{(ksec)}       &
	   \colhead{(ksec)}       &
	   \colhead{(ksec)}	  &
	   \colhead{(ksec)}         } 
\startdata 
Mrk\,335         & 00 06 19.5  & $+$20 12 10 &  0.0258 & 17.1  &  97.0 &  2.4 & 1.6 \\ 
I\,Zw1           & 00 53 34.9  & $+$12 41 36.0 & 0.0607  &\nodata&  38.6 &\nodata & \nodata  \\
Ton\,S180        & 00 57 20.0  & $-$22 22 59.3 & 0.0620  &\nodata&  16.6 & \nodata & \nodata  \\
Ton\,S210 &   01 21 51.5   & $-$28 20 57  & 0.1160    &    22.5   & 53.3  &    5.0  &  5.5   \\
Fairall\,9       & 01 23 46.0  & $-$58 48 23.8 & 0.0461  &\nodata&  38.9 & \nodata & \nodata  \\
HE\,0226$-$4410  & 02 28 15.2  & $-$40 57 16 &  0.4950    & 43.8  &  33.2 & 6.7 & 7.8 \\ 
NGC\,985         & 02 34 37.8  & $-$08 47 15.6 & 0.0431  &\nodata&  68.0 &  \nodata & \nodata  \\
PKS\,0312$-$770  & 03 11 55.2  & $-$76 51 51 &  0.2230 &  8.4  &   5.5 &  \nodata & \nodata  \\ 
PKS\,0405$-$123  & 04 07 48.4  & $-$12 11 37 &  0.5726 & 27.2  &  71.1 & 22.2 &11.1 \\ 
Akn\,120         & 05 16 11.4  & $-$00 08 59.4 & 0.0331  &\nodata&  56.2 &  \nodata & \nodata   \\
HS\,0624$+$6907  & 06 30 02.5  & $+$69 05 04 &  0.3700 & 62.0  & 112.3 & \nodata & \nodata  \\ 
VII\,Zw118       & 07 07 13.1  & $+$64 35 58.8 & 0.0797  &\nodata& 198.6 & \nodata & \nodata   \\
PG\,0804$+$761   & 08 10 58.5  & $+$76 02 41.9 & 0.1000  &\nodata& 174.0 &6.7 & 6.3 \\
Ton\,951         & 08 47 42.5  & $+$34 45 03.5 & 0.0640  &\nodata&  31.9 & \nodata & \nodata   \\
PG\,0953$+$414   & 09 56 52.4  & $+$41 15 22 &  0.2341 &  8.0  &  72.1 &  \nodata & \nodata  \\ 
Ton\,28          & 10 04 02.5  & $+$28 55 35 &  0.3297 & 33.0  &  11.2 &  \nodata & \nodata  \\ 
3C\,249.1        & 11 04 13.7  & $+$76 58 58 &  0.3115 & 68.8  & 216.8 &  \nodata & \nodata  \\ 
Mrk\,421         & 11 04 27.3  & $+$38 12 32.0 & 0.0300  &\nodata&  83.9 & 1.7 & 2.4  \\
PG\,1116$+$215   & 11 19 08.6  & $+$21 19 18 &  0.1765 & 26.5  &  77.0 &  \nodata & \nodata  \\ 
PG\,1211$+$143   & 12 14 17.7  & $+$14 03 13 &  0.0809 & 42.5  &  52.3 &  \nodata & \nodata  \\ 
PG\,1216$+$069   & 12 19 20.9  & $+$06 38 38 &  0.3313 &  5.8  &  12.0 &  \nodata & \nodata  \\ 
Mrk\,205         & 12 21 44.0  & $+$75 18 38 &  0.0708 & 62.1  & 203.6 &  \nodata & \nodata  \\ 
3C\,273          & 12 29 06.7  & $+$02 03 09 &  0.1583 & 18.7  &  42.3 &   \nodata & \nodata  \\ 
Q\,1230$+$0115   & 12 30 50.0  & $+$01 15 23 &  0.1170 &  9.8  &   4.0 & 11.1 & 8.1 \\ 
PG\,1259$+$593   & 13 01 12.9  & $+$59 02 07 &  0.4778 & 95.8  & 668.3 & 9.2 & 11.2 \\ 
PKS\,1302$-$102  & 13 05 33.0  & $-$10 33 19 &  0.2784 &  4.8  & 142.7 & 7.0 & 6.9 \\ 
Mrk\,279         & 13 53 03.4  & $+$69 18 30 &  0.0305 & 54.6  & 228.5 &  2.2 & 2.7 \\ 
NGC\,5548        & 14 17 59.5  & $+$25 08 12 &  0.0172 & 69.8  &  55.0 & 1.9 & 2.4 \\ 
Mrk\,1383        & 14 29 06.6  & $+$01 17 06 &  0.0865 & 10.5  &  63.5 &  \nodata & \nodata  \\ 
Mrk\,817         & 14 36 22.1  & $+$58 47 39.5 & 0.0313  &\nodata& 189.9 & 3.4 & 3.0  \\
Mrk\,478         & 14 42 07.5  & $+$35 26 22.9 & 0.0791  &\nodata&  14.2 &  \nodata & \nodata   \\
PG\,1444$+$407   & 14 46 45.9  & $+$40 35 06 &  0.2673 & 48.6  &  10.0 &  \nodata & \nodata  \\ 
Mrk\,290         & 15 35 52.4  & $+$57 54 09.3 & 0.0296  &\nodata&  12.8 & 3.9 & 4.8  \\
Mrk\,876         & 16 13 57.2  & $+$65 43 10 &  0.1290 & 29.2  &  46.0 & 12.6 & 11.8 \\ 
3C\,351          & 17 04 41.4  & $+$60 44 31 &  0.3719 & 77.0  & 141.9 &  \nodata & \nodata  \\ 
H\,1821$+$643    & 18 21 57.3  & $+$64 20 36 &  0.2970 & 50.9  & 132.3 & 0.6 & 0.5\\ 
PKS\,2005$-$489  & 20 09 25.4  & $-$48 49 53.9 & 0.0710  &\nodata&  49.2 & 2.5 & 1.9  \\
Mrk\,509         & 20 44 09.7  & $-$10 43 25 &  0.0344 &  7.6  &  62.3 &  9.0 & 16.5 \\ 
II\,Zw136        & 21 32 27.8  & $+$10 08 19.4 & 0.0630  &\nodata&  22.7 &  \nodata & \nodata   \\
PHL\,1811        & 21 55 01.5  & $-$09 22 25 &  0.1900 & 33.9  &  75.0 & 3.5 & 3.1 \\ 
PKS\,2155$-$304  & 21 58 52.0  & $-$30 13 32 &  0.1160 & 10.8  & 123.2 & \nodata & \nodata \\ 
Akn\,564         & 22 42 39.3  & $+$29 43 31 &  0.0247 & 10.3  &  60.9 &  1.7 & 2.4\\ 
MR\,2251$-$178   & 22 54 05.8  & $-$17 34 55.0 & 0.0644  &\nodata&  54.1 &  \nodata & \nodata  \\
NGC\,7469        & 23 03 15.6  & $+$08 52 26 &  0.0163 & 22.8  &  44.3 &  1.9 & 1.8 \\
\enddata      
\end{deluxetable*}


%% file: linestab.tex


\begin{deluxetable*}{lrlrcrrcrrrrrrrr}
\tabletypesize{\tiny}
\tablecaption{IGM absorption line properties\label{tab:linestab}}
\tablecolumns{16} 
\tablewidth{0pt} 
\tablehead{\colhead{Sight}   &  \colhead{$\lambda_{\rm obs}$}         &	   \colhead{$z$}        & 	   \colhead{Line}&	   \colhead{Ion}         &           \colhead{$W$}         &	   \colhead{$W$}     &	  \colhead{$W_{\rm MAD}$}     &	   \colhead{Ref.}             &        	   \colhead{$b$}   & \colhead{$b$}& \colhead{$b_{\rm MAD}$}& \colhead{Ref.}& \colhead{$W_{\rm min}$}& \colhead{Alt.}& \colhead{COS} \\
\colhead{Line}&\colhead{(\AA)}&&& \colhead{\#}&\colhead{Flag}&\colhead{(m\AA)}&\colhead{(m\AA)}&&\colhead{Flag}&\colhead{(km/s)}&\colhead{(km/s)}&&&\colhead{ ID}& \\
\colhead{(1) }& \colhead{(2) }& \colhead{(3) }& \colhead{(4) }& \colhead{(5) }&\colhead{ (6) }&\colhead{ (7) }& \colhead{(8) }& \colhead{(9)}& \colhead{(10)}& \colhead{(11)}& \colhead{(12)}& \colhead{(13)}& \colhead{(14)}& \colhead{(15)}& \colhead{(16)} \\} 
\startdata 

      3C249 &  1596.95 &   0.31364 &       Ly$\alpha$ &    1 &   n &$  148\pm   8$&   &        06 &   n &$  13\pm  1$&   &      23 &     21.0 &                       &    \\
      3C249 &  1591.56 &   0.30920 &       Ly$\alpha$ &    2 &   n &$   30\pm   3$&   &        02 &   n &$   8\pm  1$&   &      02 &     29.0 &                       &    \\
      3C249 &  1590.21 &   0.30809 &       Ly$\alpha$ &    3 &   n &$  262\pm  17$&   &        02 &   n &$  33\pm  1$&   &      02 &     29.0 &                       &    \\
      3C249 &  1589.95 &   0.30788 &       Ly$\alpha$ &    4 &   n &$   68\pm  19$&   &        02 &   n &$  22\pm  4$&   &      02 &     29.0 &                       &    \\
      3C249 &  1573.39 &   0.29426 &       Ly$\alpha$ &    5 &   n &$   40\pm   8$&   &        02 &   n &$  10\pm  1$&   &      02 &     28.9 &                       &    \\
      3C249 &  1544.16 &   0.27021 &       Ly$\alpha$ &    6 &   n &$   73\pm  16$&   &        02 &     {\raise.17ex\hbox{$\scriptstyle\sim$}} &$  43$&   &      02 &     47.9 &                       &    \\
      3C249 &  1539.82 &   0.26664 &       Ly$\alpha$ &    7 &   n &$   96\pm   8$&   &        02 &   n &$  67\pm 19$&   &      02 &     51.2 &                       &    \\
      3C249 &  1537.49 &   0.26473 &       Ly$\alpha$ &    8 &   n &$  210\pm  19$&   &        02 &   n &$  33\pm  2$&   &      02 &     44.5 &                       &    \\
      3C249 &  1532.51 &   0.27021 &  Si{\,III}1207 &    9 &   n &$   18\pm   6$&   &        02 &     {\raise.17ex\hbox{$\scriptstyle\sim$}} &$  19$&   &      02 &     44.2 &                       &    \\
      3C249 &  1515.64 &   0.24676 &       Ly$\alpha$ &   10 &   n &$  452\pm  16$&  17 &      0206 &   n &$  39\pm  1$&   &      02 &     51.1 &                       &    \\
      3C249 &  1512.24 &   0.24396 &       Ly$\alpha$ &   11 &   n &$  473\pm  23$&   &        02 &   n &$  56\pm  2$&   &      02 &     55.4 &                       &    \\
      3C249 &  1511.47 &   0.24332 &       Ly$\alpha$ &   12 &   n &$  317\pm  27$&   &        02 &   n &$  53\pm  3$&   &      02 &     55.4 &                       &    \\
      3C249 &  1510.34 &   0.24239 &       Ly$\alpha$ &   13 &   n &$  176\pm  20$&   &        02 &   n &$  41\pm  3$&   &      02 &     55.4 &                       &    \\
      3C249 &  1505.34 &   0.23828 &       Ly$\alpha$ &   14 &   n &$  164\pm  25$&   &        02 &   n &$  14\pm  1$&   &      02 &     56.1 &                       &    \\
      3C249 &  1504.27 &   0.24681 &  Si{\,III}1207 &   15 &   n &$   22\pm   9$&   &        23 &   n &$   7\pm  3$&   &      23 &     55.7 &                       &    \\
      3C249 &  1459.93 &   0.20093 &       Ly$\alpha$ &   16 &   n &$   78\pm  18$&   &        02 &   n &$  12\pm  2$&   &      02 &     57.2 &                       &    \\
      3C249 &  1459.63 &   0.20068 &       Ly$\alpha$ &   17 &   n &$  312\pm  11$&   &        02 &   n &$  34\pm  3$&   &      02 &     57.2 &                       &    \\
      3C249 &  1458.69 &   0.17371 &     N{\,V}1243 &   18 &   \textless &$   17$&   &        02 &     &&   &         &     58.4 &                       &    \\
      3C249 &  1455.10 &   0.19695 &       Ly$\alpha$ &   19 &   n &$   57\pm   4$&   &        02 &     {\raise.17ex\hbox{$\scriptstyle\sim$}} &$  43$&   &      02 &     42.5 &                       &    \\
      3C249 &  1454.02 &   0.17371 &     N{\,V}1239 &   18 &   n &$   29\pm   4$&   &        02 &   n &$  24\pm  7$&   &      02 &     43.3 &                       &    \\
      3C249 &  1441.77 &   0.18599 &       Ly$\alpha$ &   20 &   n &$   53\pm  10$&   &        02 &   n &$  23\pm  4$&   &      02 &     35.9 &                       &    \\
      3C249 &  1429.54 &   0.17593 &       Ly$\alpha$ &   21 &   n &$  238\pm  36$&   &        02 &   n &$  39\pm  6$&   &      02 &     49.8 &                       &    \\
      3C249 &  1429.06 &   0.17553 &       Ly$\alpha$ &   22 &   n &$  244\pm  39$&   &        02 &   n &$  41\pm  6$&   &      02 &     49.8 &                       &    \\
      3C249 &  1426.84 &   0.17371 &       Ly$\alpha$ &   23 &   n &$   49\pm  12$&   &        02 &   n &$  29\pm  5$&   &      02 &     27.0 &                       &    \\
      3C249 &  1423.54 &   0.17099 &       Ly$\alpha$ &   24 &   n &$   34\pm   10$&   &        23 &   n &$  23\pm  5$&   &      23 &     27.0 &                       &    \\
      3C249 &  1412.80 &   0.17099 &  Si{\,III}1207 &   25 &   n &$   16\pm   3$&   &        02 &   {\raise.17ex\hbox{$\scriptstyle\sim$}} &$  10$&   &      02 &     31.5 &                       &    \\
      3C249 &  1410.87 &   0.16057 &       Ly$\alpha$ &   26 &   n &$   73\pm  10$&   &        02 &   n &$  29\pm  3$&   &      02 &     31.2 &                       &    \\
      3C249 &  1396.38 &   0.24396 &  Fe{\,III}1123 &   27 &   n &$   16\pm   3$&   &        02 &   {\raise.17ex\hbox{$\scriptstyle\sim$}} &$   4$&   &      02 &     24.8 &                       &    \\
      3C249 &  1391.01 &   0.11925 &     N{\,V}1243 &   28 &   \textless &$   10$&   &        02 &     &&   &         &     32.0 &                       &    \\
      3C249 &  1388.09 &   0.14183 &       Ly$\alpha$ &   29 &   n &$  278\pm  10$&   &        02 &   n &$  38\pm  1$&   &      02 &     31.3 &                       &    \\
      3C249 &  1386.55 &   0.11925 &     N{\,V}1239 &   28 &   n &$   11\pm   2$&   &        02 &   {\raise.17ex\hbox{$\scriptstyle\sim$}} &$  15$&   &      02 &     32.5 &                       &    \\
      3C249 &  1384.73 &   0.13907 &       Ly$\alpha$ &   30 &   n &$   27\pm   9$&   &        02 &   n &$  25\pm  7$&   &      02 &     31.9 &                       &    \\
      3C249 &  1379.42 &   0.13470 &       Ly$\alpha$ &   31 &   n &$   64\pm   3$&   &        02 &   n &$  65\pm  2$&   &      02 &     25.1 &                       &    \\
      3C249 &  1368.50 &   0.12572 &       Ly$\alpha$ &   32 &   n &$  373\pm  21$&   &        02 &   n &$  31\pm  1$&   &      02 &     28.7 &                       &    \\
      3C249 &  1368.07 &   0.12536 &       Ly$\alpha$ &   33 &   n &$  151\pm  21$&   &        02 &   n &$  43\pm  5$&   &      02 &     28.7 &                       &    \\
      3C249 &  1363.05 &   0.31364 &    O{\,VI}1038 &   34 &   n &$   22\pm  11$&   &        06 &   n &$  14\pm  2$&   &      23 &     24.7 &                       &    \\

\enddata      
\tablecomments{Table 4 is published in its entirety in the electronic edition of the Astrophysical Journal. A portion is shown here for guidance regarding its form and content. A complete description of the columns in this table can be found in Section~\ref{sec:construct} of the text.}
\end{deluxetable*}

%% file: ionstab.tex

\begin{deluxetable*}{lrlccrrcrr}
\tabletypesize{\tiny}
\tablecaption{IGM absorber properties\label{tab:ionstab}}
\tablecolumns{10} 
\tablewidth{0pt} 
\tablehead{\colhead{Sight Line}   &           \colhead{$z$}         &	   \colhead{Ion}        & 	   \colhead{Ion code}&	   \colhead{$N$ Flag}         &          \colhead{$\log N$}         &	   \colhead{Ref.}     &	   \colhead{$b$ Flag}     &	   \colhead{$b$}             &        	   \colhead{Ref.}    \\
&&&&&\colhead{ ($N$ in \unit{\centi\meter\rpsquared}) }&&& \colhead{(\unit{\kilo\meter\usk\reciprocal\second})}& \\
\colhead{(1) }& \colhead{(2) }& \colhead{(3) }& \colhead{(4) }& \colhead{(5) }&\colhead{ (6) }&\colhead{ (7) }& \colhead{(8) }& \colhead{(9)}& \colhead{(10)}\\      } 
\startdata 
      3C249 &   0.31364 &      H\,I &    1 &   n &$ 13.976\pm 0.021$&        1623 &   n &$  12.5\pm   0.6$&      1623 \\
      3C249 &   0.30920 &      H\,I &    2 &   n &$ 12.720\pm 0.040$&          02 &   n &$   8.0\pm   1.0$&        02 \\
      3C249 &   0.30809 &      H\,I &    3 &   n &$ 13.860\pm 0.085$&          02 &   n &$  35.7\pm   5.6$&        02 \\
      3C249 &   0.30788 &      H\,I &    4 &   n &$ 13.050\pm 0.140$&          02 &   n &$  22.0\pm   4.0$&        02 \\
      3C249 &   0.29426 &      H\,I &    5 &   n &$ 12.860\pm 0.080$&          02 &   n &$  10.0\pm   1.0$&        02 \\
      3C249 &   0.27021 &      H\,I &    6 &   n &$ 13.210\pm 0.125$&          02 &   {\raise.17ex\hbox{$\scriptstyle\sim$}}&$  43$&        02 \\
      3C249 &   0.26664 &      H\,I &    7 &   n &$ 13.320\pm 0.075$&          02 &   n &$  67.0\pm  19.0$&        02 \\
      3C249 &   0.26473 &      H\,I &    8 &   n &$ 13.900\pm 0.100$&          02 &   n &$  27.5\pm   4.5$&        02 \\
      3C249 &   0.27021 &   Si\,III &    9 &   n &$ 12.030\pm 0.190$&          02 &  {\raise.17ex\hbox{$\scriptstyle\sim$}} &$  19$&        02 \\
      3C249 &   0.24676 &      H\,I &   10 &   n &$ 14.460\pm 0.067$&        0203 &   n &$  37.2\pm   1.2$&      0203 \\
      3C249 &   0.24396 &      H\,I &   11 &   n &$ 14.200\pm 0.035$&          02 &   n &$  56.9\pm   4.7$&        02 \\
      3C249 &   0.24332 &      H\,I &   12 &   n &$ 13.940\pm 0.080$&          02 &   n &$  52.8\pm   8.7$&        02 \\
      3C249 &   0.24239 &      H\,I &   13 &   n &$ 13.800\pm 0.055$&          02 &   n &$  30.7\pm   3.8$&        02 \\
      3C249 &   0.23828 &      H\,I &   14 &   n &$ 13.840\pm 0.120$&          02 &   n &$  14.0\pm   1.0$&        02 \\
      3C249 &   0.24681 &   Si\,III &   15 &   n &$ 12.099\pm 0.125$&        0323 &   n &$   8.4\pm   3.2$&      0323 \\
      3C249 &   0.20093 &      H\,I &   16 &   n &$ 13.250\pm 0.100$&          02 &   n &$  12.0\pm   2.0$&        02 \\
      3C249 &   0.20068 &      H\,I &   17 &   n &$ 13.970\pm 0.090$&          02 &   n &$  38.5\pm   8.9$&        02 \\
      3C249 &   0.17371 &      NV &   18 &   n &$ 13.190\pm 0.130$&          02 &   n &$  24.0\pm   7.0$&        02 \\
      3C249 &   0.19695 &      H\,I &   19 &   n &$ 13.080\pm 0.090$&          02 &   {\raise.17ex\hbox{$\scriptstyle\sim$}} &$  43$&        02 \\
      3C249 &   0.18599 &      H\,I &   20 &   n &$ 12.970\pm 0.070$&          02 &   n &$  23.0\pm   4.0$&        02 \\
      3C249 &   0.17593 &      H\,I &   21 &   n &$ 13.740\pm 0.060$&          02 &   n &$  39.0\pm   6.0$&        02 \\
\enddata      
\tablecomments{Table 5 is published in its entirety in the electronic edition of the Astrophysical Journal. A portion is shown here for guidance regarding its form and content. A complete description of the columns in this table can be found in Section~\ref{sec:construct} of the text.}
\end{deluxetable*}


%% file: tablepaper.tex
\begin{deluxetable*}{lrrlrrrrr}
\tabletypesize{\tiny}
\tablecaption{Summary of IGM Detections and Results\label{tab:detect}}
\tablecolumns{9} 
\tablewidth{0pt} 
\tablehead{
\colhead{Ion }& \colhead{$\mathcal{N}_{\rm total}$ }& \colhead{$\mathcal{N}_{\rm}$ }& \colhead{$z_{\rm abs}$ }& \colhead{$\Delta z_{\rm max}$ }& \colhead{$d\mathcal{N}/dz$\tablenotemark{a} }& \colhead{$d\mathcal{N}/dz$\tablenotemark{a} }&\colhead{ $d\mathcal{N}/dz$\tablenotemark{a} }& \colhead{$\beta$} \\
 &  &  &  & &  \colhead{($>\unit{10}{\milli\angstrom}$) }&\colhead{($>\unit{21}{\milli\angstrom}$) }&\colhead{  ($>\unit{30}{\milli\angstrom}$)} &  \\
\colhead{(1) }& \colhead{(2) }& \colhead{(3) }& \colhead{(4) }& \colhead{(5) }&\colhead{ (6) }&\colhead{ (7) }& \colhead{(8) }& \colhead{(9)} \\
    } 
\startdata

O\,VI & 118 & 111 & $<0.40$ & 6.095 & $42.0^{+10.7}_{ -6.3}$ & $27.7^{+  3.8}_{ -2.9}$ &$  22.2^{+  3.2}_{ -2.4}$ & $2.075\pm0.119$\\
N\,V & 29 & 25 & $<0.396 $& 5.408 & $70.4^{+ 47.1}_{-24.1}$ & $20.6^{+  8.2 }_{-5.1}$ & $9.6^{+  5.4}_{ -2.7}$ & $2.001\pm0.235$\\
C\,IV & 36 & 29 & $<0.116 $& 2.523 & $14.3^{+  4.5}_{ -2.5}$ & $14.3^{+  4.5}_{ -2.5}$ & $11.7^{+  3.7}_{ -2.0}$ & $1.774\pm0.157$\\
C\,III & 51 & 50 & $<0.40 $& 5.644 & $23.8^{+ 10.9}_{ -4.7}$ & $17.3^{+  4.7}_{ -2.7}$ & $14.4^{+  3.7}_{ -2.2}$ & $1.859\pm0.092$\\
Si\,IV & 34 & 30 & $<0.24$ & 4.317 & $15.4^{+  6.7}_{ -3.0}$ & $11.8^{+  4.0}_{ -2.1}$ & $9.2^{+  3.5}_{ -1.8}$ & $1.729\pm0.125$\\
Si\,III & 62 & 57 & $<0.40$& 5.244 & $16.6^{+  4.1}_{ -2.6}$ & $10.4^{+  2.4 }_{-1.5}$ & $7.4^{+  2.1}_{ -1.2}$ & $1.739\pm0.102$\\
H\,I & 797 & 746 & $<0.40$ & 5.382 & $283.9^{+ 57.2}_{-32.8}$ & $175.2^{+ 13.0 }_{-9.8}$ & $144.0^{+  6.6 }_{-6.0}$ & $1.680\pm0.030$\\
\cutinhead{Poorly Constrained Ion Species}

Fe\,III & 15 & 11 & $<0.40$ & 5.853 & $9.3^{+  5.2}_{ -2.6}$ &  $5.4^{+  4.2 }_{-1.8}$ & $3.7^{+  3.5}_{ -1.4}$ & $2.216\pm0.428$\\
C\,II & 18 & 18 & $<0.296$ & 4.900 & $9.6^{+5.3}_{-2.2}$ & $8.0^{+3.7}_{-1.8}$ & $7.5^{+3.5}_{-1.7}$ & $1.587\pm0.121$\\
Fe\,II & 5 & 4 & $<0.40$ & 5.913 & $11.2^{+13.8}_{-5.0}$ & $4.3^{+7.0}_{-2.5}$ & $4.3^{+7.0}_{-2.5}$ & $1.922\pm1.505$\\
Si\,II & 15 & 15 & $<0.372$ &5.345& $7.1^{+3.6}_{ -1.7}$ &  $7.1^{+3.6}_{-1.7}$ & $6.1^{+3.3}_{ -1.5}$ & $1.568\pm0.157$\\
S\,II & 3 & 3 & $<0.373$ &5.352& $10.0^{+13.4}_{ -4.8}$ &  $6.2^{+10.1}_{-3.6}$ & $3.1^{+7.1}_{ -2.5}$ & $1.568\pm0.157$\\

S\,III\tablenotemark{b} & 1 & 1 & $<0.40$ &5.924&   \nodata &  \nodata& \nodata & \nodata\\
S\,IV\tablenotemark{b} & 0 & 0 & $<0.40$ &5.889&   \nodata &  \nodata& \nodata & \nodata\\

N\,II\tablenotemark{c} & 8 & 8 & \nodata &\nodata& \nodata &  \nodata& \nodata & \nodata\\
N\,III\tablenotemark{c} & 3 & 3 & \nodata &\nodata& \nodata &  \nodata& \nodata & \nodata

\enddata      
\tablenotetext{a}{Number of absorbers per unit redshift path length, integrated down to the indicated equivalent width.}
\tablenotetext{b}{No statistics are presented for this ion species owing to a lack of detections.}
\tablenotetext{c}{Though several detections of this ion are reported in the catalog, no systematic search was made for this species. As a result, all detections were serendipitous and any statistics derived from those detections may be heavily biased. We therefore refrain from reporting any such statistics but note the detections for completeness.}
\end{deluxetable*}

%% file: sec243.tex
\subsubsection{Note on Inconsistencies in DS08}

The calculations of contributions to $\Omega$ by both metals and \ion{H}{1} were described and calculated inconsistently 
in \citetalias{dan08} owing to incorrect cosmological factors. 
The baryon fractions listed in 
their Tables 12 and 13 were computed with incorrect cosmological corrections that differed from those given in Equations 7--10 of \citetalias{dan08} and consequently overestimated values of 
$\Omega_b$ for all ions discussed.  Our new \textit{HST} archive survey properly includes these effects, as well as the redshift evolution 
of the hydrogen photoionization rate,  $\Gamma_H(z)$.  The original baryon fraction values tabulated in \citetalias{dan08} are unsuitable 
for comparison to other observations or simulations, though the absorber statistics that do not depend on $dX$ (e.g., $d\mathcal{N}/dz$) remain valid. Detailed corrections of the data tables will be submitted as an erratum.
For the sake of comparison to the present work, we briefly describe the issue here. 

\citetalias{dan08} calculated contributions to $\Omega_b$ using a redshift-invariant path length $dX$ instead of $dz$, but
the definition of  $dX$ was applied inconsistently. The text described the conventional ``absorption pathlength function'' 
of \citet{bah69}:
\begin{equation}
       dX \equiv (1+z)^2~[\Omega_m(1+z)^3+\Omega_\Lambda]^{-1/2}~dz  \; , 
\end{equation}
but the calculations incorrrectly implemented the pathlength function as
\begin{equation}
       dX \equiv (1+z)^{-1}~[\Omega_m(1+z)^3+\Omega_\Lambda]^{-1/2}~dz  \;  .  
\end{equation}
The difference in the two formalisms is a factor of $(1+z)^3$, negligible at $z\sim0$ but a factor of $\sim3$ at the highest redshifts 
surveyed.  As a result, the corrected $d{\cal N}/dX$ values using the proper formalism as presented in the erratum are
lower than those published in Table~11 of \citetalias{dan08} by $25-39\%$ for most ion species.  Because \ion{C}{4} is only seen 
 at $z\la0.1$ in the STIS/E140M data, its value of $d{\cal N}/dX$ only decreases by roughly 13\%.

The cosmological parameters $\Omega_{\rm ion}$ and $\Omega_{\rm IGM}^{\rm (ion)}$ (\citetalias{dan08}, Equations~6 and 8 
and Table~12) scale with $d{\cal N}/dX$.  Recalculating the values of Table~12 with the correct $dX$ treatment results in a $\sim40\%$ decrease in the values of $\Omega_{\rm IGM}^{\rm (ion)}$ in most cases ($\sim12\%$ for 
\ion{C}{4}).  Interestingly, this $\sim40\%$ decrease in $\Omega_{\rm IGM}^{\rm (OVI)}/\Omega_b$ brings the DS08 value more 
in line with other published values at the same sensitivity limits.  

The overall baryon density accounted for by Ly$\alpha$ absorbers (Equations~9 and 10 of \citetalias{dan08}) suffer similar 
adjustments, changing to $\Omega^{\rm (HI)}_{\rm IGM}/\Omega_b=20.7\pm2.4\%$ if calculated with the \citet{pen04} method
and $\Omega^{\rm (HI)}_{\rm IGM}/\Omega_b=17.5\pm2.6\%$ if calculated with the \citet{sch01} method over a column density range, $12.5<\log N_{\rm HI}<16.5$. These adjusted values still do not account for the redshift evolution of $\Gamma_H (z)\propto (1+z)^{4.4}$ which is included in the method of \citet{shu12} as applied to the present paper's catalog and enters the equation as the square root.

All further references in the present paper to values of $\Omega_b$ reported by \citetalias{dan08} refer to the 
corrected values.

%% file: omega_lya_allmethod.tex
%
%
\begin{deluxetable*}{lc||cc||cc||cc}
\tabletypesize{\tiny}
\tablecolumns{8} 
\tablewidth{0pt} 
\tablecaption{Baryon Content of the Local Ly$\alpha$ Forest}
\tablehead{\colhead{$\log\,N_{\rm HI}$ Range}    &
           \colhead{$\cal N$}     &
           \colhead{$\Omega_{\rm Ly\alpha}$\,\tablenotemark{a}}     &
           \colhead{$\Omega_{\rm Ly\alpha}/\Omega_b~(\%)$\,\tablenotemark{a}}   &
           \colhead{$\Omega_{\rm Ly\alpha}$\,\tablenotemark{b}}     &
           \colhead{$\Omega_{\rm Ly\alpha}/\Omega_b~(\%)$\,\tablenotemark{b}}   &
           \colhead{$\Omega_{\rm Ly\alpha}$\,\tablenotemark{c}}     &
           \colhead{$\Omega_{\rm Ly\alpha}/\Omega_b~(\%)$\,\tablenotemark{c}}   }
\startdata
$ 12.0-12.5 $&$  20 $&$ 0.0055\pm0.0043 $&$ 12.0\pm 9.5 $&$ 0.0038\pm0.0041 $&$  8.3\pm 9.0 $&$ 0.0036\pm0.0041 $&$  8.0\pm 8.9 $\\
$ 12.5-13.5 $&$ 427 $&$ 0.0044\pm0.0010 $&$  9.8\pm 2.2 $&$ 0.0039\pm0.0008 $&$  8.6\pm 1.8 $&$ 0.0038\pm0.0008 $&$  8.4\pm 1.7 $\\
$ 13.5-14.5 $&$ 244 $&$ 0.0031\pm0.0003 $&$  6.7\pm 0.6 $&$ 0.0038\pm0.0003 $&$  8.4\pm 0.8 $&$ 0.0038\pm0.0003 $&$  8.3\pm 0.7 $\\
$ 14.5-15.5 $&$  49 $&$ 0.0012\pm0.0002 $&$  2.7\pm 0.5 $&$ 0.0023\pm0.0004 $&$  5.0\pm 1.0 $&$ 0.0022\pm0.0004 $&$  4.9\pm 0.9 $\\
$ 15.5-16.5 $&$   3 $&$ 0.0001\pm0.0001 $&$  0.3\pm 0.2 $&$ 0.0003\pm0.0002 $&$  0.7\pm 0.4 $&$ 0.0003\pm0.0002 $&$  0.7\pm 0.4 $\\
&&&&&&\\
$ 12.5-14.5 $&$ 671 $&$ 0.0075\pm0.0010 $&$ 16.5\pm 2.3 $&$ 0.0077\pm0.0009 $&$ 17.0\pm 1.9 $&$ 0.0076\pm0.0008 $&$ 16.6\pm 1.8 $\\
$ 14.5-16.5 $&$  52 $&$ 0.0013\pm0.0002 $&$  2.9\pm 0.5 $&$ 0.0026\pm0.0005 $&$  5.7\pm 1.0 $&$ 0.0026\pm0.0005 $&$  5.7\pm 1.0 $\\
$ 16.5-19.0 $&$   3 $&$ 0.0006\pm0.0004 $&$  1.4\pm 0.8 $&$ 0.0035\pm0.0022 $&$  7.8\pm 4.8 $&$ 0.0035\pm0.0021 $&$  7.7\pm 4.7 $\\
&&&&&&&\\
$ 12.5-16.5 $&$ 723 $&$ 0.0088\pm0.0011 $&$ 19.4\pm 2.3 $&$ 0.0103\pm0.0010 $&$ 22.7\pm 2.2 $&${ 0.0102\pm0.0010} $&$ 22.3\pm 2.1 $\\
\hline
&&\\
$ 12.5-19.0 $&$ 726   $&\multicolumn{4}{c}{Shull method for $\log\,N_{\rm HI}<16.5$, Schaye method for $\log\,N_{\rm HI}>16.5$:}&$ 0.0108\pm0.0010 $&$ 23.7\pm 2.2   $
\enddata
\tablenotetext{a}{Method of \citet{pen00}}
\tablenotetext{b}{Method of \citet{sch01}}
\tablenotetext{c}{Method of \citet{shu12}\bigskip}
  \label{tab:omega_lya}
\end{deluxetable*}

%% file: tablepaperomega.tex
\begin{deluxetable*}{lrrlr}
\tabletypesize{\small}
\tablecaption{Summary of IGM Baryon Fraction Results\label{tab:omega}}
\tablecolumns{5} 
\tablewidth{0pt} 
\tablehead{\colhead{Ion }&\multicolumn{2}{c}{$\Omega_{\rm ion}~(10^{-8})$} &\multicolumn{2}{c}{$\Omega_{\rm IGM}^{\rm (ion)}/\Omega_b$\tablenotemark{a}}  \\
 &   \colhead{($>\unit{10}{\milli\angstrom}$) }&\colhead{  ($>\unit{30}{\milli\angstrom}$)}  &  \colhead{($>\unit{10}{\milli\angstrom}$) }&\colhead{  ($>\unit{30}{\milli\angstrom}$)}   } 
\startdata

\ion{O}{6}&$48.22\pm3.77$&$40.33\pm4.28$&$0.086\pm0.007$&$0.072\pm0.008$\\
\ion{N}{5}&$4.19\pm2.25$&$2.42\pm2.85$&$0.059\pm0.032$&$0.034\pm0.040$\\
\ion{C}{4}&$20.11\pm2.01$&$19.83\pm2.06$&$0.067\pm0.007$&$0.066\pm0.007$\\
\ion{C}{3}&$10.83\pm1.38$&$8.75\pm1.46$&$0.013\pm0.002$&$0.010\pm0.002$\\
\ion{Si}{4}&$11.07\pm1.57$&$10.78\pm1.78$&$0.099\pm0.014$&$0.097\pm0.016$\\
\ion{Si}{3}&$17.53\pm3.21$&$17.26\pm4.27$&$0.061\pm0.011$&$0.060\pm0.015$\\
\ion{Fe}{3}&$40.16\pm14.31$&$29.48\pm16.93$&$0.082\pm0.029$&$0.060\pm0.035$
\enddata      
\tablenotetext{a}{{Scaled by $f_{\rm ion}$, $Z$, and $H_0$; $\Omega_b=0.0455\pm0.0028$ \citep{kom11}. See text for details.\bigskip}}
\end{deluxetable*}